\documentclass[11pt]{iopart}
\usepackage{graphicx}

\usepackage{csquotes}
\usepackage{lipsum}
\usepackage{float}
\usepackage{xcolor}
\usepackage{subcaption}
\usepackage{gensymb}
\usepackage{multirow}

\usepackage[backend=biber,style=authortitle,bibstyle=numeric, citestyle=phys,doi=false,url=false,isbn=false, maxnames = 20]{biblatex}

\begin{document}


\title[Deformable Scintillation Dosimeter I : Challenges and correction techniques]{Deformable Scintillation Dosimeter I: Challenges and Implementation using Computer Vision Techniques}




\author{E Cloutier$^{1,2}$, L Archambault$^{1,2}$,
L Beaulieu$^{1,2}$\footnote{Present address:
Department of Physics, Université Laval, Quebec,
QC BS8 1TS, Canada.}
\address{$^1$ Physics, physical engineering and optics department and Cancer Research Center, Universite Laval, Quebec, Canada.}
\address{$^2$ CHU de Quebec – Université Laval, CHU de Quebec, Quebec, Canada}}

\begin{abstract}
Plastic scintillation detectors are increasingly used to measure dose distributions in the context of radiotherapy treatments. Their water-equivalence, real-time response and high spatial resolution distinguish them from traditional detectors, especially in complex irradiation geometries. Their range of applications could be further extended by embedding scintillators in a deformable matrix  mimicking anatomical changes. In this work, we characterized signal variations arising from the translation and rotation of scintillating fibers with respect to a camera. Corrections are proposed using stereo vision techniques and two sCMOS complementing a CCD camera. The study was extended to the case of a prototype real-time deformable dosimeter comprising an array of 19 scintillating. The signal to angle relationship follows a gaussian distribution (FWHM = 52$\degree$) whereas the intensity variation from radial displacement follows the inverse square law. Tracking the position and angle of the fibers enabled the correction of these spatial dependencies. The detecting system provides an accuracy and precision of respectively 0.008 cm and 0.03 cm on the position detection. This resulted in an uncertainty of  2$\degree$ on the angle measurement. Displacing the dosimeter by $\pm 3$ cm in depth resulted in relative intensities of $100\pm10\%$ (mean $\pm$ standard deviation) to the reference position. Applying corrections reduced the variations thus resulting in relative intensities of $100\pm1\%$. Similarly, for lateral displacements of $\pm 3$ cm, intensities went from $98\pm3\%$ to $100\pm1\%$ after the correction.  Therefore, accurate correction of the signal collected by a camera imaging the output of scintillating elements in a 3D volume is possible. This work paves the way to the development of real-time scintillator-based deformable dosimeters.

\end{abstract}

\maketitle


\section{Introduction}

Over the last decade, water-equivalent radio-luminescent materials have been used in a variety of setups to quantify delivered dose distributions of radiotherapy treatments. From plastic scintillating fiber detectors to volumetric scintillation dosimeters and Cherenkov imaging, such systems enable real-time measurements with high spatial resolution over a wide range of energies \cite{pogue_cherenkov_2015,  beaulieu_review_2016, beddar_water_2006}, without the need for energy-dependent correction factors. Moreover, the advent of complex personalized treatment plans using a greater number of small fields, more modulated beams and magnetic fields \cite{alexander_1_nodate, madden_first_2019, therriault-proulx_effect_2018} highlight the advantages of plastic scintillation detectors making them well suited tools for the rising challenges of advanced radiation therapy techniques \cite{beddar_scintillation_2016}.  

Over the same period, a growing clinical interest to consider anatomical variations in treatment planning and delivery has developed.  Inter-fractional and intra-fractional organ motion, as well as anatomical deformations, have been shown to result in clinically significant dose variations that need to be accounted for \cite{schaly_tracking_2004, brock_inclusion_2003}. This adds another layer of complexity for dose measurements. Therefore, there is an increasing need for new dosimeters capable of measuring dose in a deformable matrix mimicking anatomical variations \cite{kirby_need_2013}. Scintillators have been used for 3D dosimetry and may be an ideal choice for measurement in the presence of deformations. Volumetric scintillation dosimeters have demonstrated the ability to perform millimeter resolution, real-time and water-equivalent dosimetry of dynamic treatment plan over 2D \cite{goulet_novel_2014} and 3D volumes \cite{rilling_simulating_nodate, rilling_tomographicbased_2020, guillot_performance_2013, kroll_preliminary_2013, kirov_three-dimensional_2005, kirov_new_2000}. A scintillator-based deformable dosimeter would be suited to the challenges imposed by both motion management and advanced radiotherapy modalities. Furthermore, given the rapidly increasing role of artificial intelligence \cite{thompson_artificial_2018} in radiation oncology, the need for accurate experimental validation will likely increase in the future. However, going from static to deformable geometries entails new difficulties. Applying a deformation to a radioluminescent-based phantom will lead to translations and rotations of the radioluminescent elements resulting in variations of the signal collected, even if no change in deposited dose is expected.

This work is the first to investigate the signal variations arising from the displacement and rotations of a point-like scintillator directly imaged by a camera (i.e. not coupled to a clear optical fiber). Using computer vision techniques, the position of the tip of a scintillating fiber and it's angulation in regards to the photo-detector is tracked, and signal variations are corrected. Then, those correction techniques are applied to the case of a deformable phantom comprising an array of 19 scintillating fibers measuring the dose from a linac. The dosimeter and correction method were subsequently applied to the simultaneous deformation vector fields and dose distribution measurements, which is presented in the companion paper \cite{part2-submittedpaper}.

\section{Methods}

Measurements were conducted with different detection setups which are summarized in table \ref{tab:summary}. Throughout this work, green scintillators (length: 1.2~cm, diameter: 0.1~cm, BCF-60; Saint-Gobain Crystals, Hiram, OH, USA) are used. All irradiations were performed with a 6 MV photon beam (Clinac iX, Varian, Palo Alto, USA).

\begin{table}[]
\caption{Summarized detection setup for each experiment. F/M indicate whether the detector and the dosimeter are fixed (F) or moving (M).}
\begin{tabular}{llclc}
\br
Experiment & Detector(s) & \begin{tabular}[c]{@{}l@{}}F/M \end{tabular} & Dosimeter& \begin{tabular}[c]{@{}l@{}}F/M\end{tabular} \\ 
\mr
\textbf{\begin{tabular}[c]{@{}l@{}}Signal caracterization \\ 
(section 2.1)\end{tabular}} && &  & \\
Angular correction & \multirow{2}{*}{\begin{tabular}[c]{@{}l@{}} sCMOS mounted\\on robot\end{tabular}} & \multirow{2}{*}{M}  & \multirow{2}{*}{\begin{tabular}[c]{@{}l@{}}One scintillator \\ at isocenter \end{tabular}} & \multirow{2}{*}{F}  \\
Distal correction  &   &  &  &  
\\
\mr
\textbf{\begin{tabular}[c]{@{}l@{}}Signal correction\\ (section 2.2)\end{tabular}}& & &  &  \\
3D positionning accuracy & sCMOS1 + CCD & \multirow{2}{*}{F}  & \multirow{2}{*}{\begin{tabular}[c]{@{}l@{}}One scintillator \\ mounted on robot\end{tabular}}  & \multirow{2}{*}{M}   \\
Angular measurement   & sCMOS2 + CCD   &    &   &  \\
\mr
\textbf{\begin{tabular}[c]{@{}l@{}}Correction validation \\ (section 2.3)\end{tabular}}  &  &  &  &  \\
Single scintillator  & \begin{tabular}[c]{@{}l@{}}sCMOS mounted\\ on robot\end{tabular}  & M  & \begin{tabular}[c]{@{}l@{}}One scintillator\\ at isocenter\end{tabular}  & F   \\
19 scintillators dosimeter & \begin{tabular}[c]{@{}l@{}}sCMOS1 + sCMOS2\\  + CCD\end{tabular}   & F   & \begin{tabular}[c]{@{}l@{}} Deformable \\ dosimeter\end{tabular}  & M  \\                   \br 
\end{tabular}
\label{tab:summary}
\end{table}

\subsection{Characterizing signal spatial dependencies}
\begin{figure}
    \centering
    \includegraphics[width = 0.7\textwidth]{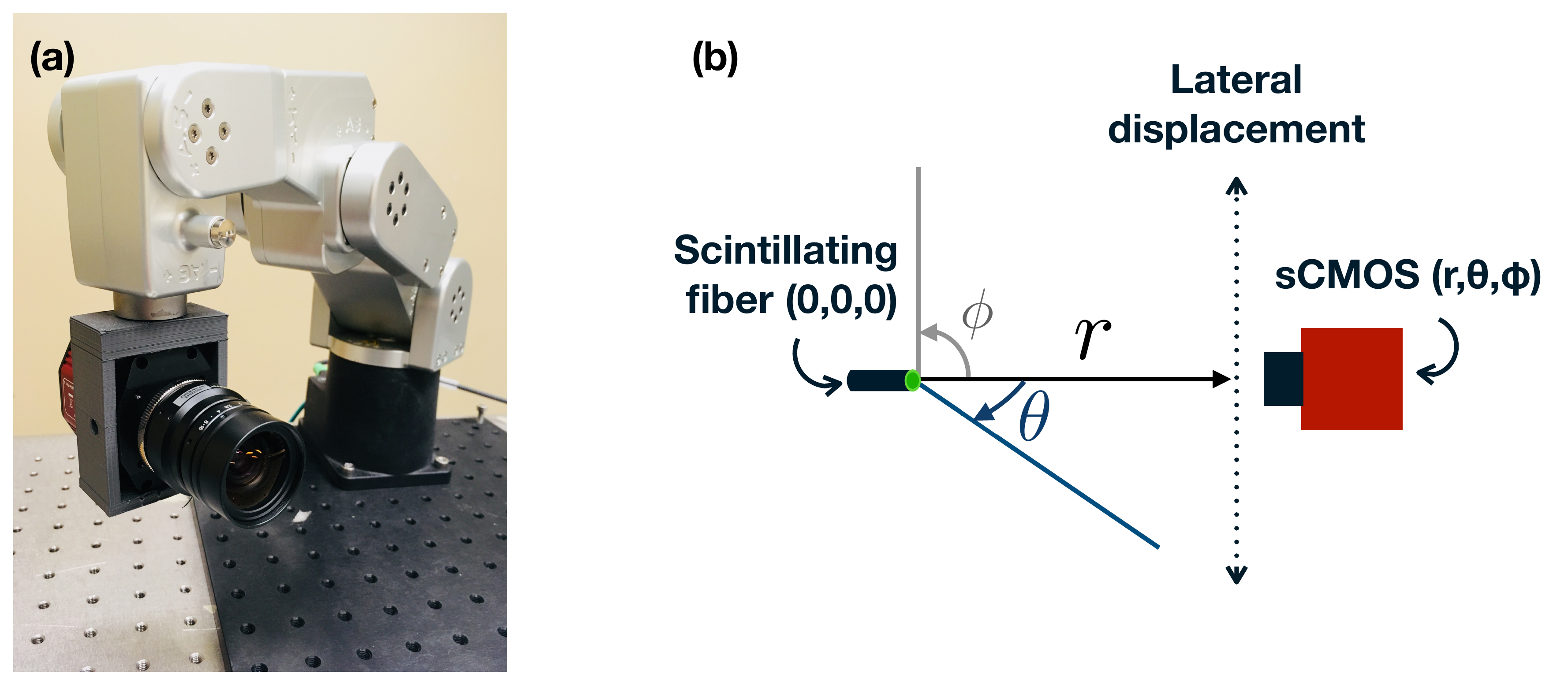}
    \caption{Setup used for the $\theta$ and $r$ calibration and validation using lateral displacements : (a) pictures the camera mounted on the robot while (b) presents the coordinate system.}
    \label{fig:methods}
\end{figure}

 Signal variations caused by the displacement and rotation of scintillating fibers were separately characterized using a sCMOS camera (Quantalux, Thorlabs, Newton, USA) mounted  on a Meca500 small industrial robot arm (Mecademic, Montreal, Canada) (figure \ref{fig:methods}). From different viewpoints, the sCMOS acquired the scintillating signal from the tip of a scintillator positioned at the isocenter of a 6 MV photon beam. All measurements were compared to the signal obtained at a reference position set to (r,~$\theta$,~$\phi$)~=~(35,~0,~0). The relation between the collected light and the orientation of the camera with respect to the scintillator was characterized by moving the camera with the robot around the scintillator, within the robot's limits ($\pm$26$\degree$), keeping a constant radial distance (r =~35 cm,~$\theta$,~$\phi$~=~0). Then, the signal to radial distance ($r$) relationship was measured by moving the camera towards to scintillating fiber, from 30 to 43 cm, keeping the orientation fixed (r,~$\theta$~=~0,~$\phi$~=~0). Acquisitions from a uniform white emitter screen were also performed to quantify the impact of vignetting in the resulting images. The vignetting for each pixel (i, j) was calculated using a $\cos^4 (\theta_{(i,j)} )$ fit as suggested by Robertson et al \cite{robertson_optical_2014}. 
 {}
\subsection{Signal corrections}

A setup of 3 cameras was designed to measure the signal, orientation and 3D position of irradiated scintillation fibers (figure \ref{fig:methods-camera}). The setup comprises two sCMOS and one cooled CCD (Alta U2000, Andor Technology, Belfast, United Kingdom). The CCD camera was chosen for its capacity to provide stable measurements, whereas the sCMOS were selected for their high spatial resolution (1920 x 1080 pixels). The resulting detection assembly aims at correcting the signal from moving scintillators measured with static cameras. 

 \subsubsection{Rotation measurement}
 
 To account for the rotation of a scintillating fibers, a sCMOS camera was positioned in front of a CCD.  Angles were calculated from the measured vertical ($dy_i$) and lateral ($dx_i$) displacement shifts by the facing cameras : 
 \begin{equation}
     \sin\theta_m = \frac{dx_1 + dx_2}{l}; \qquad \sin\phi_m = \frac{dy_2 - dy_1}{l}.
 \end{equation}
 The accuracy of tilt measurements was assessed by mounting a 1.2 cm length (0.1 cm diameter) scintillating fiber on the robot arm. $\theta_m$ and $\phi_m$ were simultaneously measured while rotating the fiber with the robot in the $\theta_r$ and $\phi_r$ direction from 0$\degree$ to 30$\degree$.

 \subsubsection{3D position tracking}

 Distance corrections rely on the 3D distance ($r^2~=~x^2~+~y^2~+~z^2$) of the fiber in the object space with respect to the camera's sensor center. Hence, a stereoscopic pair of camera was used to project the 2D image position of each scintillating fiber onto the 3D object space and correct variations resulting from changes in their optical coupling with the cameras. Using computer vision, it is possible to project the 3D object position $\underline{\underline{\widetilde{P}}}(x, y, z)$ on a 2D image plane $\underline{\widetilde{p}}(x', y')$ using a projective transformation as:
\begin{equation}
  \underline{\widetilde{p}} = s\underline{\widetilde{m}} = \underline{\underline{K}}_{3\times 3}\left.[\underline{\underline{R}}^t -\underline{\underline{R}}^t t]\right._{3\times 4} \underline{\underline{\widetilde{P}}}.
\end{equation}
$\underline{\underline{K}}$ and $[\underline{\underline{R}}^t -\underline{\underline{R}}^t t]$ respectively refer to the intrinsic and extrinsic parameters of the camera, which can be extracted from calibration \cite{hartley_multiple_2004}. The intrinsic parameters matrix depends on the properties of the detector used whereas the extrinsic parameters matrix depends on the position (rotation, translation) of the detectors with regards to the imaged scene. Once known, it is possible to reconstruct the (x,y) position of an image point in the object space. However, using only one camera limits the projection to (x,y) coordinates as the z position (depth) is degenerate. The use of an additional camera imaging the object from a different perspective removes the degeneracy along z and enables the 3D positioning of the object.

 \begin{figure}
    \centering
    \includegraphics[width = 0.8\textwidth]{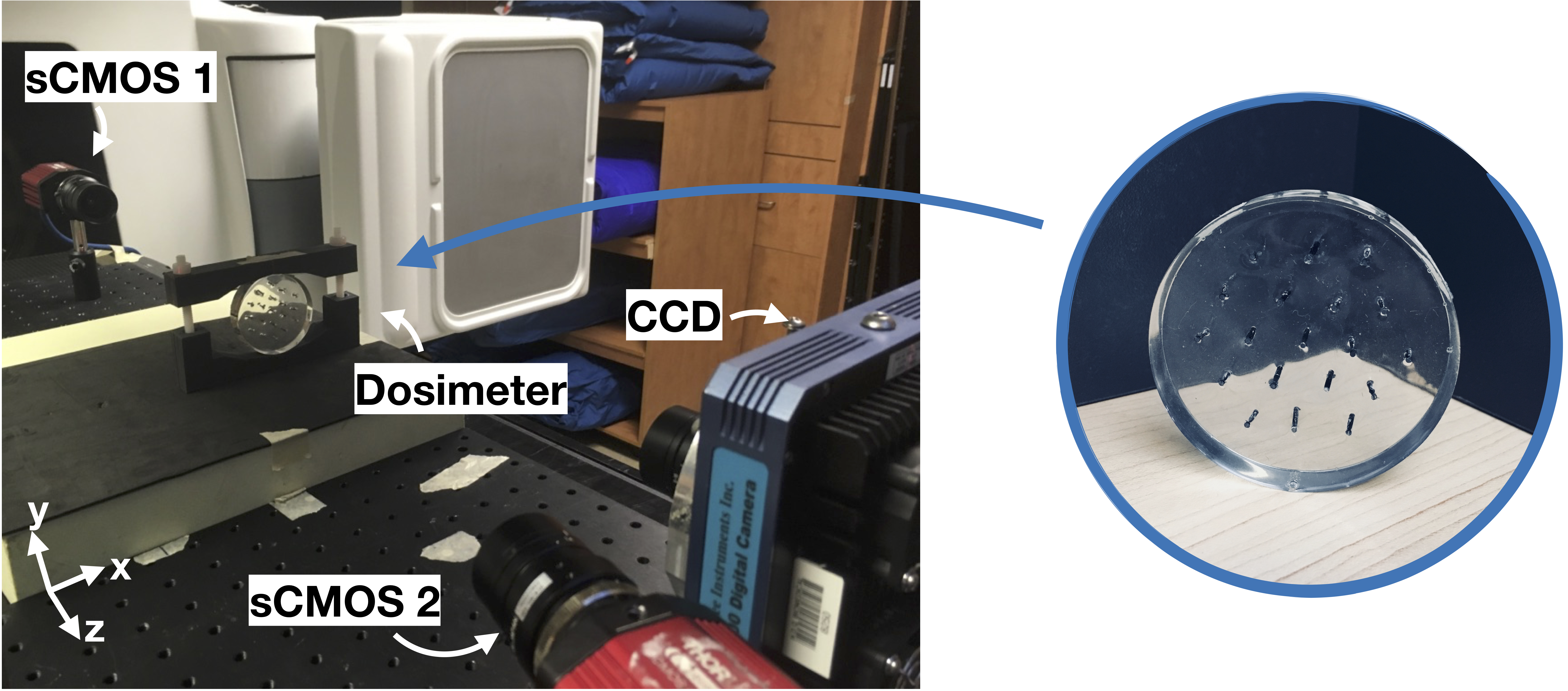}
    \caption{Illustration of the room set-up for irradiation measurements: the CCD measuring the scintillation signal and the two sCMOS paired to correct the signal from variations in the distance and orientation of the fibers. The sCMOS1 is used to measure the tilts of the scintillating fibers while the sCMOS2 enables their 3D tracking. The dosimeter comprises an array of 19 scintillating fibers distant by 1 cm.}
    \label{fig:methods-camera}
\end{figure}{}

In this work, we paired the Alta U2000 cooled CCD to a sCMOS to locate the tip scintillating fibers in the object space $(x, y, z)$. With this location, it was possible to apply the distance correction to signal variations arising from the movement of the fibers. Cameras were calibrated using a (15$\times$10) chessboard pattern and a calibration algorithm inspired by Zhang from the OpenCV Python library version 3.4.2 \cite{zhang_flexible_2000, opencv_library}. Images were rectified and corrected for distortion before performing the triangulation. The rectification eases triangulation calculations whereas the distortion correction increases its accuracy. The position of the left camera in relation to the first one, as obtained from calibration, is presented in table \ref{tab:rot_cam}. The accuracy of the 3D tracking from stereo-vision was assessed by mounting a scintillating fiber on the robot arm. Displacements in the x, y and z axis were subsequently performed by the robot in increments of 1 cm.

\begin{table}
\caption{Translation and rotation between the sCMOS$_2$ and the CCD coordinate systems as obtained from the calibration}
\centering
\begin{tabular}{@{}l|lll}
\br
Translation [cm] & X : 10.67 & Y : -0.33 & Z : 3.45\\
\mr
Angle [$\degree$] & Pitch : -1.3 & Yaw : -15.62 & Roll : -0.26 \\
\br
\end{tabular} \\

\label{tab:rot_cam}
\end{table}

 \subsection{Validation of signal correction}
 
The signal resulting from the lateral displacement was acquired to validate the proposed correction technique. The case of a single scintillating fiber was first assessed by mounting a sCMOS on the robot imaging a fixed scintillating fiber. Measurement were taken from -7 to 7 cm in increments of 1 mm and a correction was performed using known distance ($r$) and orientation ($\theta$,~$\phi$).  

\subsection{Application to a deformable scintillating detector}
The method was extended to the case of a deformable scintillator-based dosimeter comprising as array of 19 BCF-60 scintillating fibers (figure \ref{fig:methods-camera}) and a complete correction was carried out without prior knowledge on the distance and orientation of the fibers.

\subsubsection{Dosimeter fabrication}
The deformable dosimeter prototype consists of a clear, flexible cylindrical elastomer in which 19 scintillating fibers were embedded (figure \ref{fig:methods-camera}). The cylinder is made from a commercial urethane liquid elastomer compound (Smooth-On, Macongie, USA) cast in a silicone cylindrical mold (diameter: 6 cm, thickness: 1.2 cm). The compound was degassed in vacuum prior to pouring in order to remove trapped air bubbles which would have reduced the final transparency of the elastomer. Nineteen scintillations fibers were inserted in the cylindrical elastomer  guided by a  3D-printed template. Each scintillating fiber was covered by a heat-shrinking opaque cladding to isolate the scintillation light from its surrounding and, more importantly, limit the collected signal to the one emerging from its ends. The scintillating fibers were embedded in the phantom forming a 1x1x1 cm triangular grid array. 

\subsubsection{Dosimeter characterization}
 The density (in g/cm$^3$) of the detector was extracted from a CT-scan (Siemens Somatom Definition AS Open 64, Siemens Healthcare, Forchheim, Germany). CT-scans of the bulk elastomer (i.e. no fibers embedded) and a reference water volume were also acquired for comparison. The pitch, current and energy of the scanner were respectively set to 0.35, 60 mA and 120 kVp. The detector was also irradiated with a 6 MV, 600 cGy/min photon beam (Clinac iX, Varian, Palo Alto, USA) while being imaged. The center of the detector was aligned with the isocenter of the linac. Dose linearity was studied while varying the dose deposited or the dose rate. Different dose rates were achieved by varying the distance between the detector and the irradiation source while keeping the integration time and delivered monitor units constant. 
 
\subsubsection{Dose correction measurements}
The dosimeter was displaced laterally from -3 to 3 cm relative to it's initial position relative to the camera. Radial displacement were also conducted moving the dosimeter from 32 to 38 cm from the CCD camera. Displacements were achieved by translating the treatment couch in 1 cm increments and repositionning the center of the dosimeter at the isocenter of the linac, to keep the dose constant. The irradiations were of 100 monitor units (MU). The radiometry, i.e. quantitative measurement of scintillating signal related to the dose, was carried by the CCD camera, while both sCMOS measured the angle and 3D position of the fibers.  The CCD camera was positioned 35 cm from the dosimeter and coupled to a 12 mm focal length lens (F/\# = 16). For each measurement, five backgrounds, i.e. images acquired in the absence of ionizing radiation, and five signal frames were acquired. Background frames were subtracted from the signal images. Then, acquisitions were corrected using a median temporal filter \cite{archambault_transient_2008} and combined with an average. The scintillation signal was integrated on the resulting images over a 3$\times$3 pixels region-of-interest centered on the centroid of each scintillating fiber. The cameras were shielded with lead blocks to reduce noise from stray radiation.

\section{Results}

\begin{figure}[ht]
    \centering
    \begin{subfigure}{.35\textwidth}
    \centering
    \includegraphics[width =0.99\textwidth ]{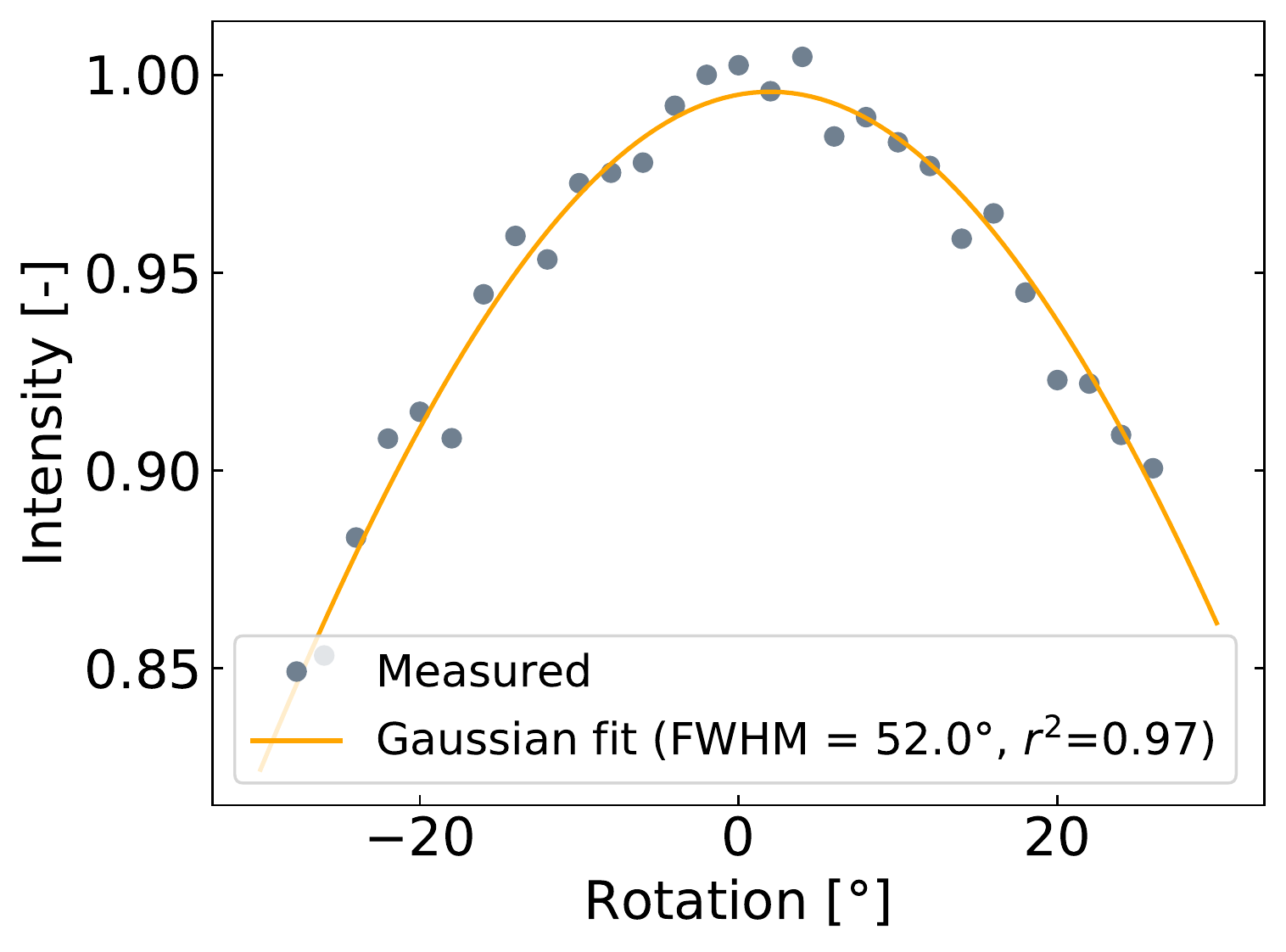}
    \caption{}
    \end{subfigure}
     \begin{subfigure}{.34\textwidth}
    \centering
    \includegraphics[width =0.99\textwidth ]{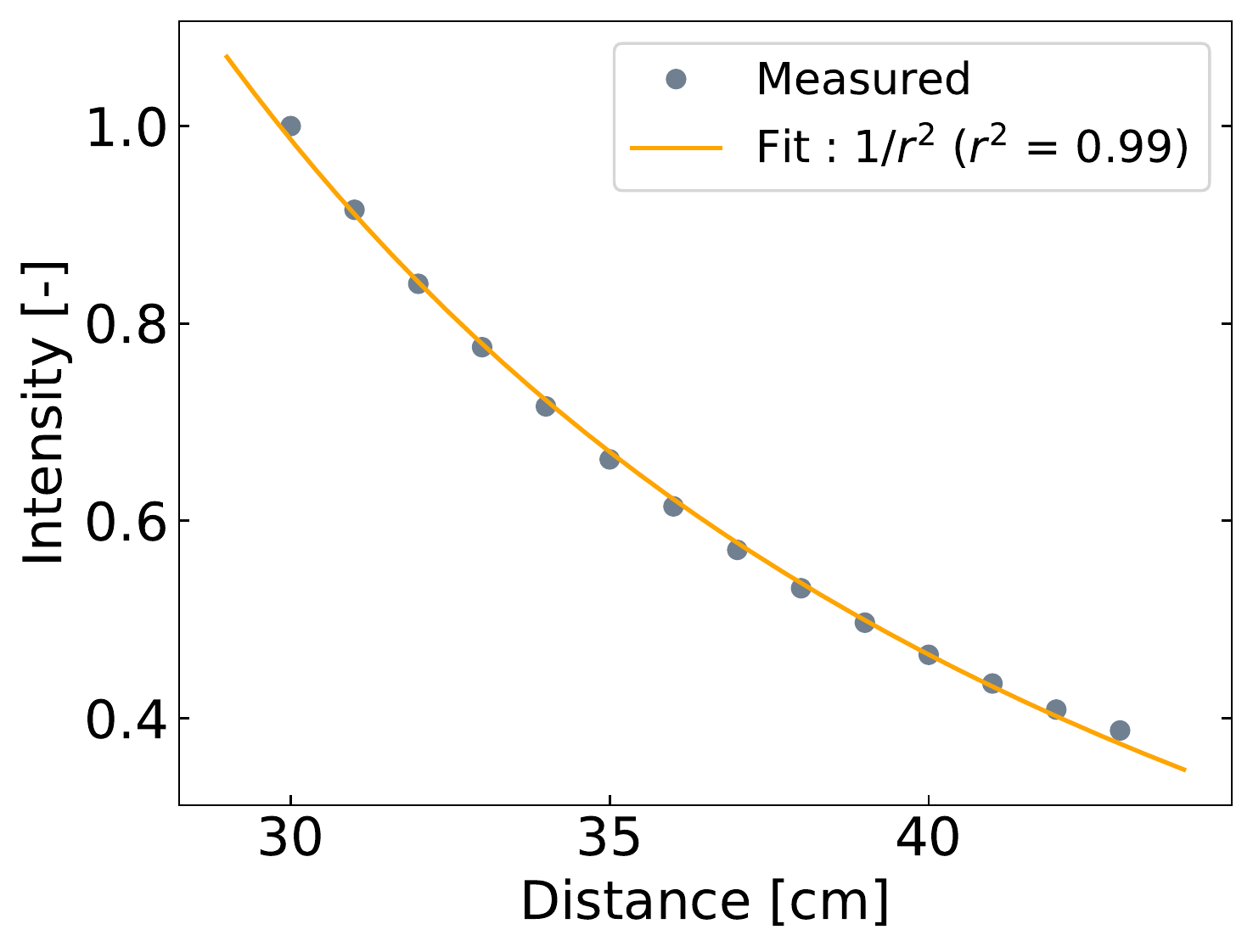}
    \caption{}
    \end{subfigure}
     \begin{subfigure}{.29\textwidth}
    \centering
    \includegraphics[width =0.99\textwidth ]{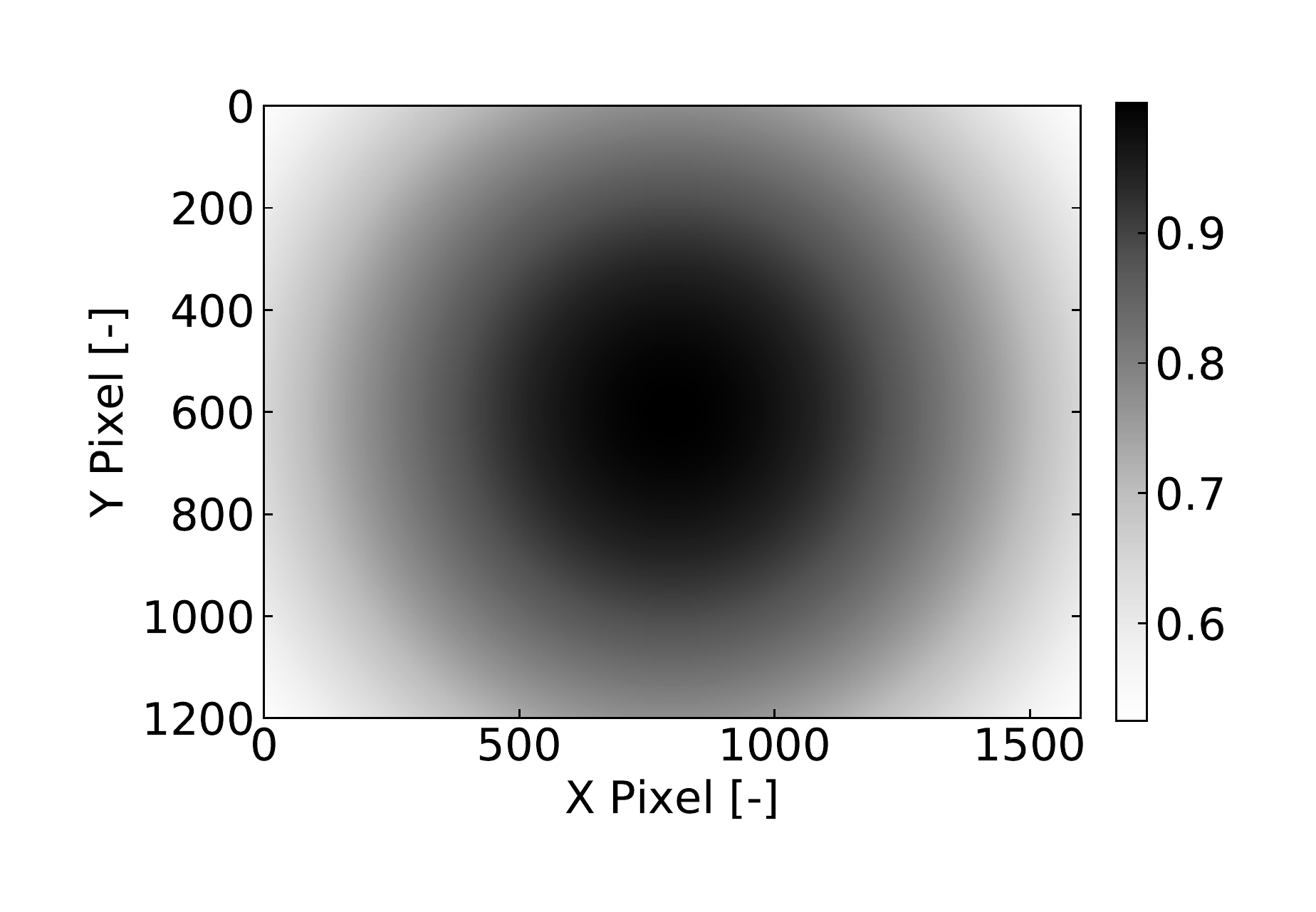}
    \caption{}
    \end{subfigure}
    \caption{Characterization of the angular $\theta$ (a) and radial $r$ (b) dependencies of the signal collected by the camera. Vignetting (c) was also characterized to correct signal variations of the sensor.}
    \label{fig:corrsignal}
\end{figure}

\subsection{Spatial dependencies characterization}
Rotating the camera's optical axis with respect to the  scintillating fibers axis results in a decrease of the collected signal. This decrease can be modeled according to a gaussian distribution with a full width at half max (FWHM) of 52 $\degree$ (see figure \ref{fig:corrsignal}a). For comparison, the scintillating fiber has a numerical aperture of 0.583 which results in an emission angle limited to 35.45$\degree$ in air. Figure \ref{fig:corrsignal}b presents the distance to signal relationship obtained while varying the distance between the camera and the scintillating fiber from 30 to 43 cm. Increasing the distance results in a decrease of the collected signal following the inverse square law ($R^2>0.99$). Finally, figure \ref{fig:corrsignal}c presents the vignetting function used to correct signal variations arising from the displacements of the scintillating spots on the CCD sensor.

\subsection{Angle measurement accuracy}

\begin{figure}[ht]
    \centering
    \begin{subfigure}{.49\textwidth}
    \centering
    \includegraphics[width =0.75\textwidth ]{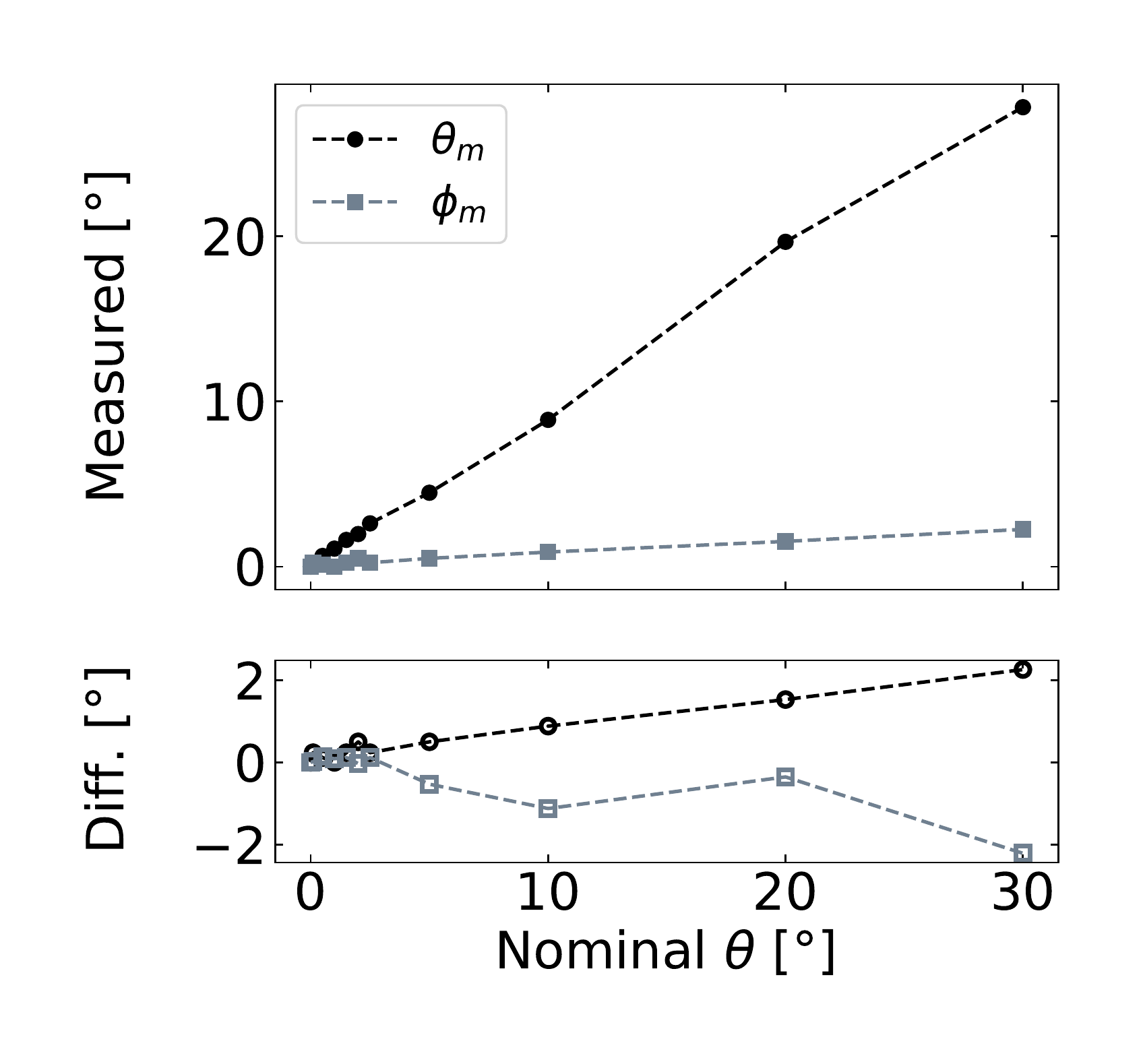}
    \caption{}
    \end{subfigure}
     \begin{subfigure}{.49\textwidth}
    \centering
    \includegraphics[width =0.75\textwidth ]{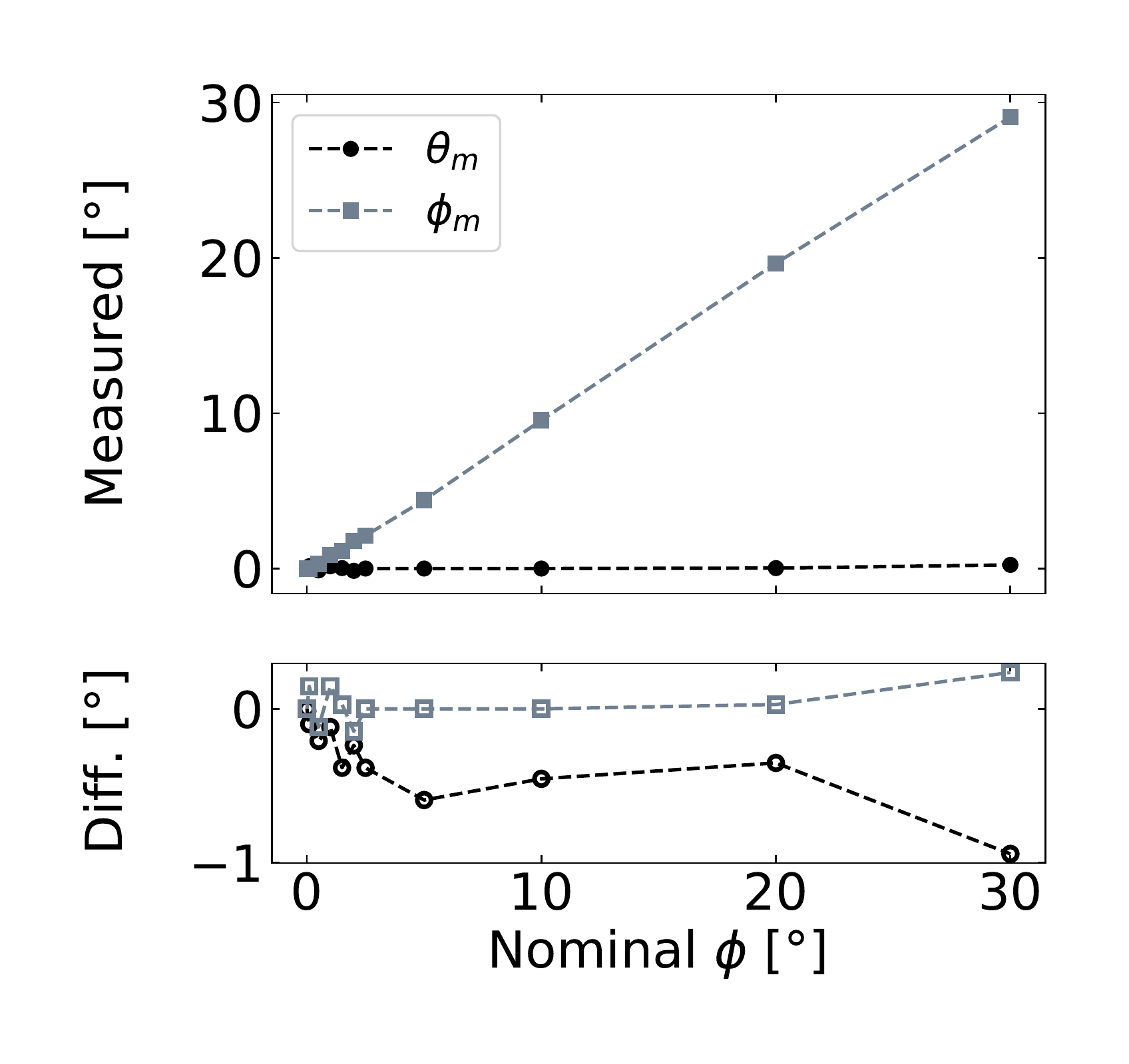}
    \caption{}
    \end{subfigure}
    \caption{$\theta_m$ and $\phi_m$ measured while rotating a scintillating fiber either in the $\theta$ (a) or $\phi$ (b) axis.}
    \label{fig:carac-corr-angle}
\end{figure}

Figure \ref{fig:carac-corr-angle} presents $\theta_m$ and $\phi_m$ measured while moving a scintillating fiber in the $\theta$ (a) and $\phi$ (b) direction. Rotating the scintillating fiber from 0 to 30$\degree$ resulted in differences up to 2.3$\degree$ between the measured and predicted tilts. 

\subsection{3D tracking accuracy}
\begin{figure}[ht]
    \centering
    \begin{subfigure}{.325\textwidth}
    \centering
    \includegraphics[width =0.99\textwidth ]{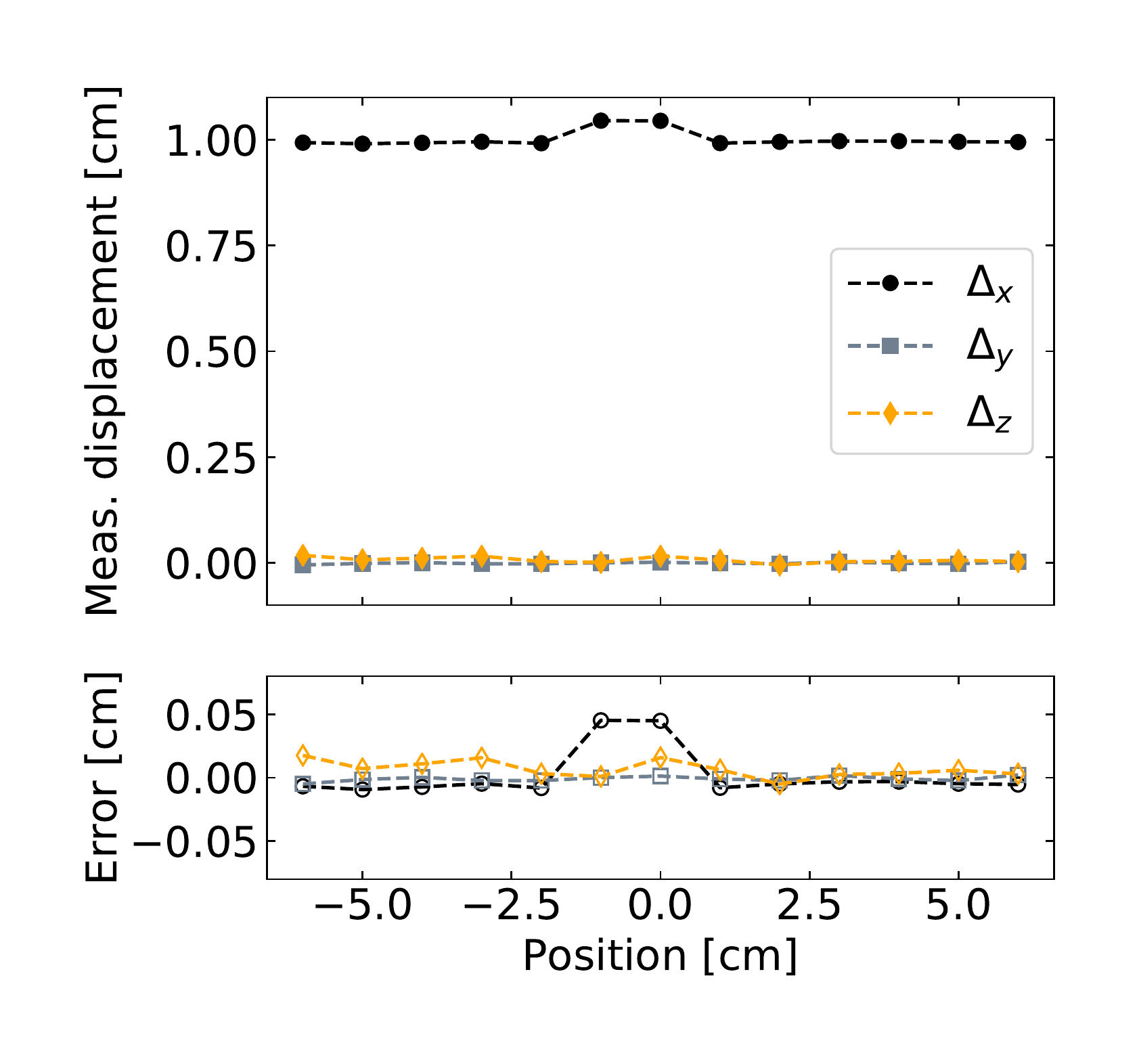}
    \caption{}
    \end{subfigure}
     \begin{subfigure}{.325\textwidth}
    \centering
    \includegraphics[width =0.99\textwidth ]{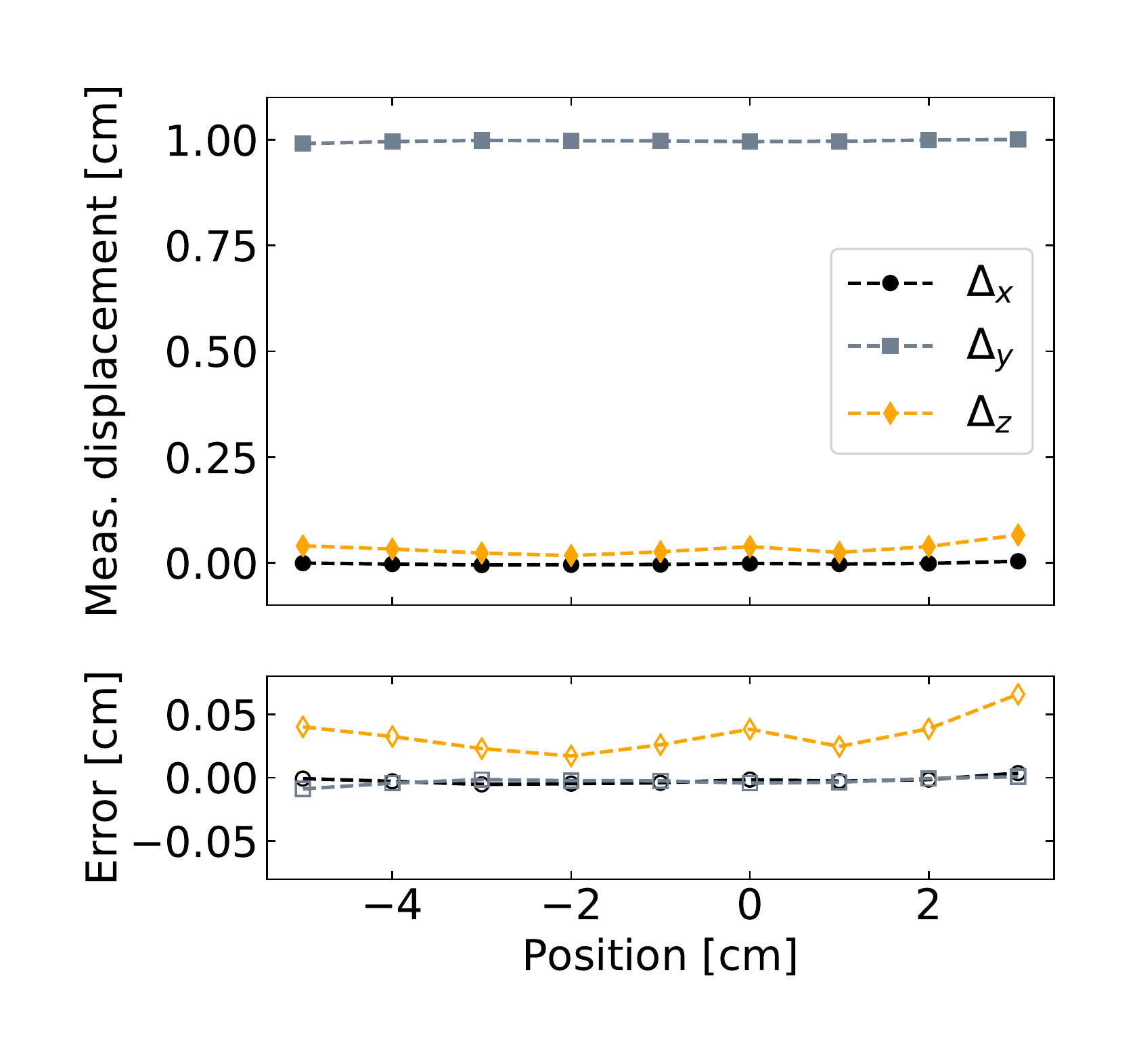}
    \caption{}
    \end{subfigure}
     \begin{subfigure}{.325\textwidth}
    \centering
    \includegraphics[width =0.99\textwidth ]{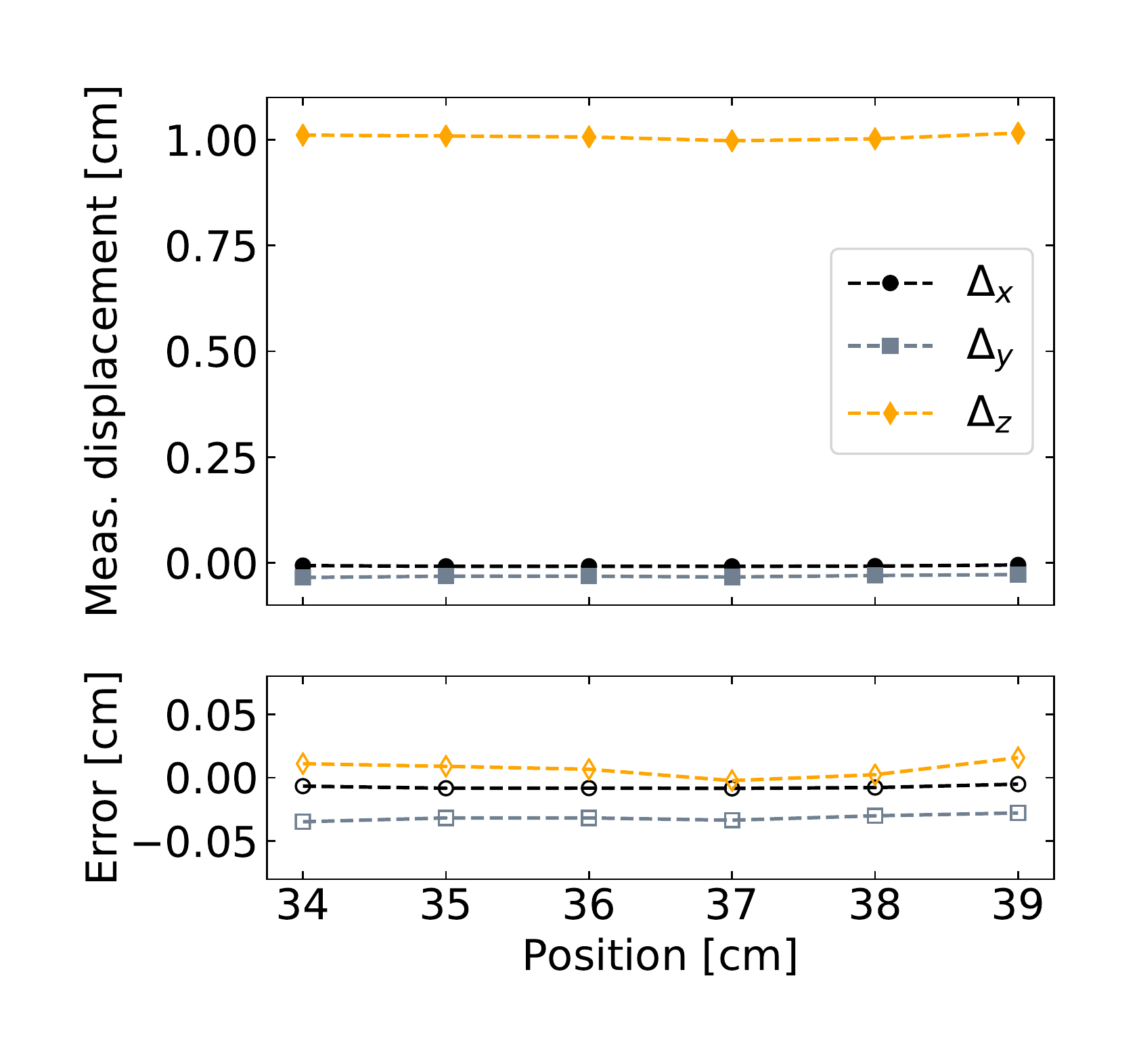}
    \caption{}
    \end{subfigure}
    \caption{3D measured displacement, while moving the scintillating fiber in the X (a), Y (b) or (c) direction. The errors represent the difference between the expected displacement and the one measured with the stereo pair of cameras.}
    \label{fig:stereo}
\end{figure}

Figure \ref{fig:stereo} presents the measured displacement by the stereoscopic pair in the x, y and z axis while moving the fibers in 1 cm increments in each directions, successively. Throughout all the measurements, the system provided a mean accuracy and precision of 0.008 cm and 0.03 cm respectively.

\subsection{Corrections validation}
\begin{figure}[ht]
    \centering
    \begin{subfigure}{.49\textwidth}
    \centering
    \includegraphics[width =0.75\textwidth ]{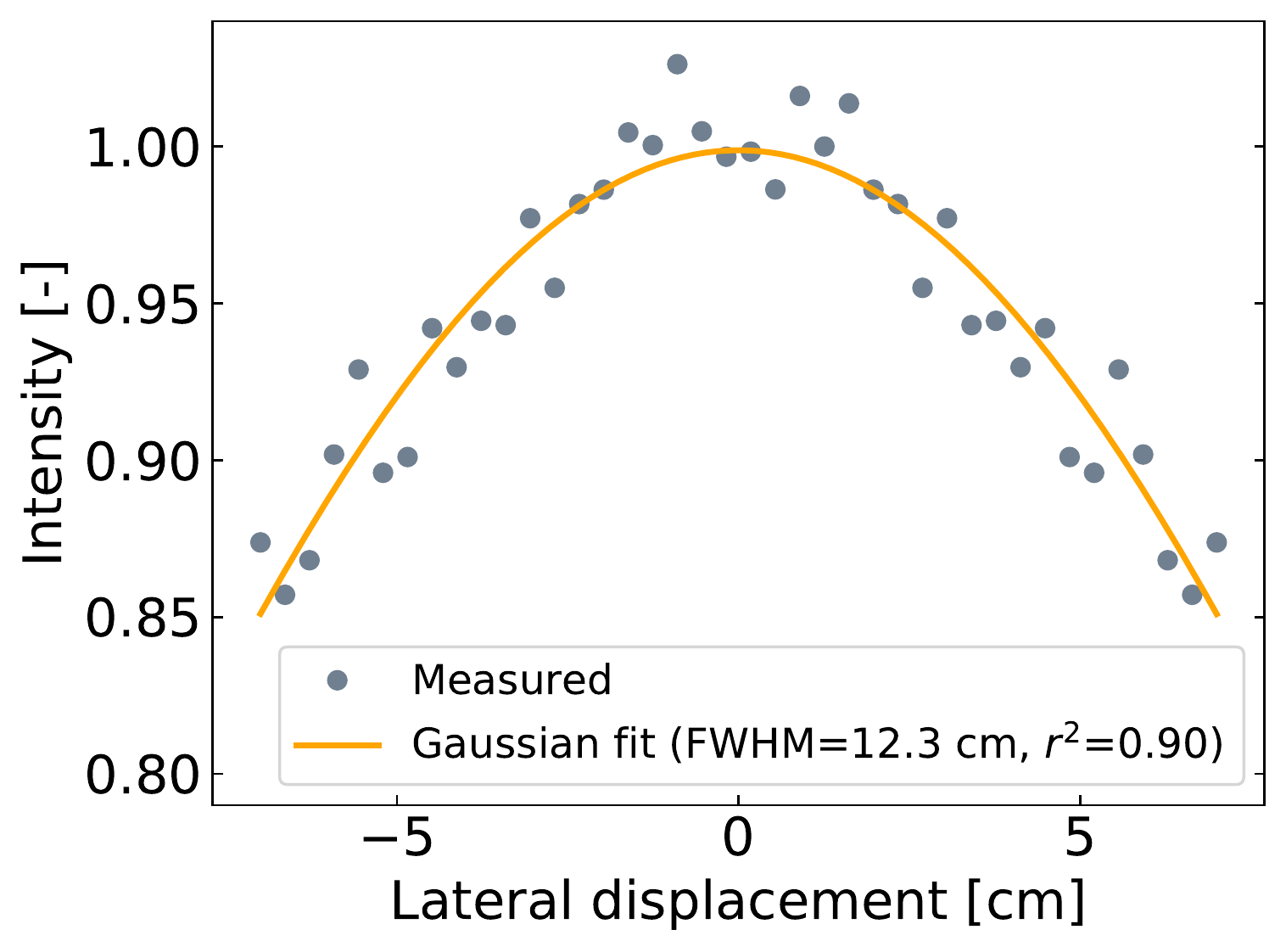}
    \caption{}
    \end{subfigure}
     \begin{subfigure}{.49\textwidth}
    \centering
    \includegraphics[width =0.75\textwidth ]{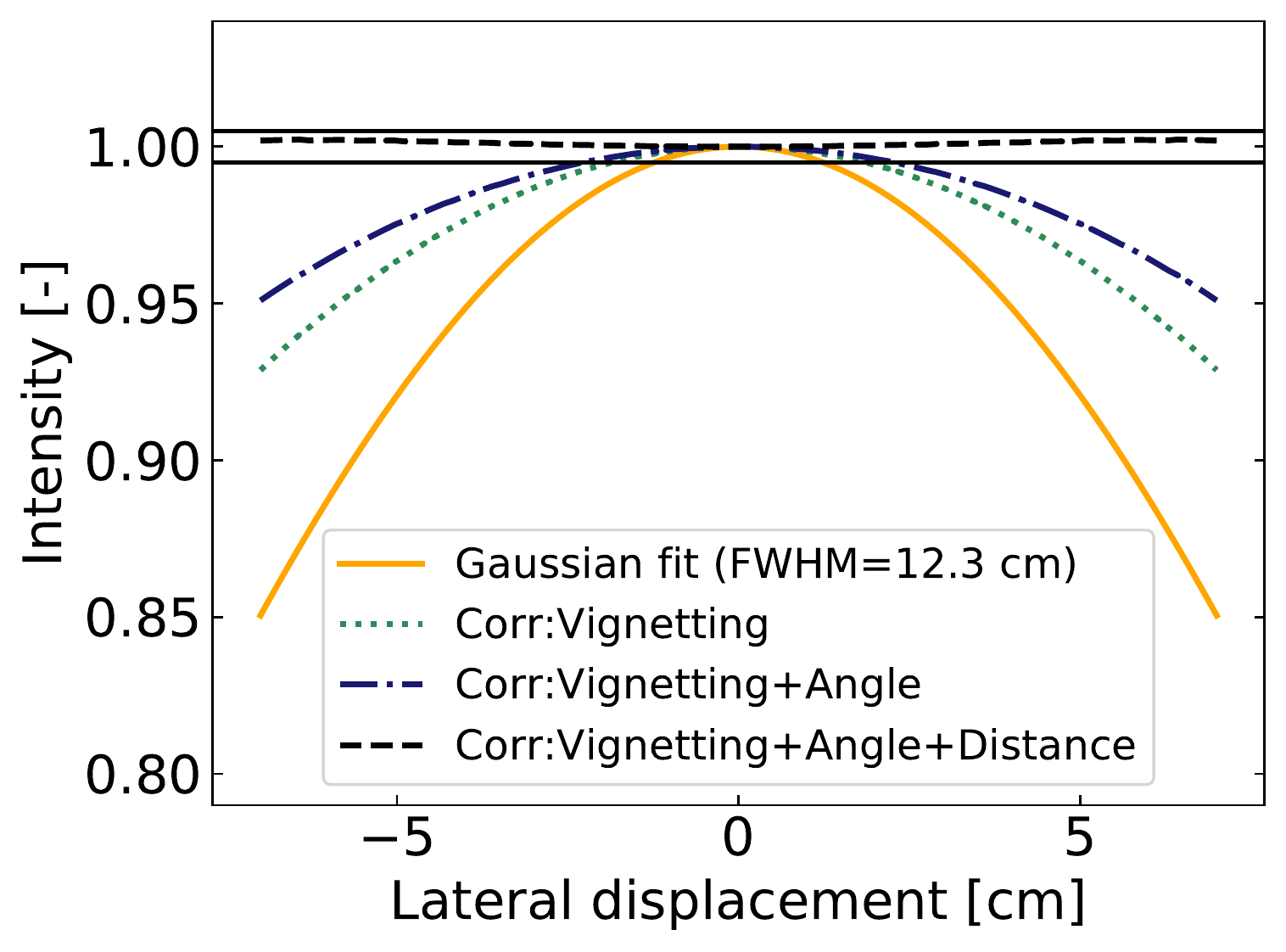}
    \caption{}
    \end{subfigure}
    \caption{Signal variation arising from lateral displacement of the camera (a) and its correction using angular, radial and vignetting corrections (b). Horizontal lines in (b) represents a $\pm$ 0.5\% variation.}
    \label{fig:latsignal}
\end{figure}

Figure \ref{fig:latsignal} presents the signal variations measured from a lateral displacement of the camera imaging one scintillating fiber and the resulting signal after angular, radial and vignetting corrections are applied. A gaussian fit was applied to raw data to account for uncertainties arising from wobbling movements of the camera through the robot's  displacement. Figure \ref{fig:latsignal}(b) shows the corrected signal and the contribution of vignetting, angle and distance to the magnitude of the correction. The combined correction resulted in signal variations lesser than 0.5\% for lateral displacements ranging from -7 to 7~cm. 

\subsection{Application to a deformable scintillating detector}
\subsubsection{Dosimeter characterization}
\begin{figure}[ht]
    \centering
    \begin{subfigure}{.45\textwidth}
    \centering
    \includegraphics[width =0.6\textwidth ]{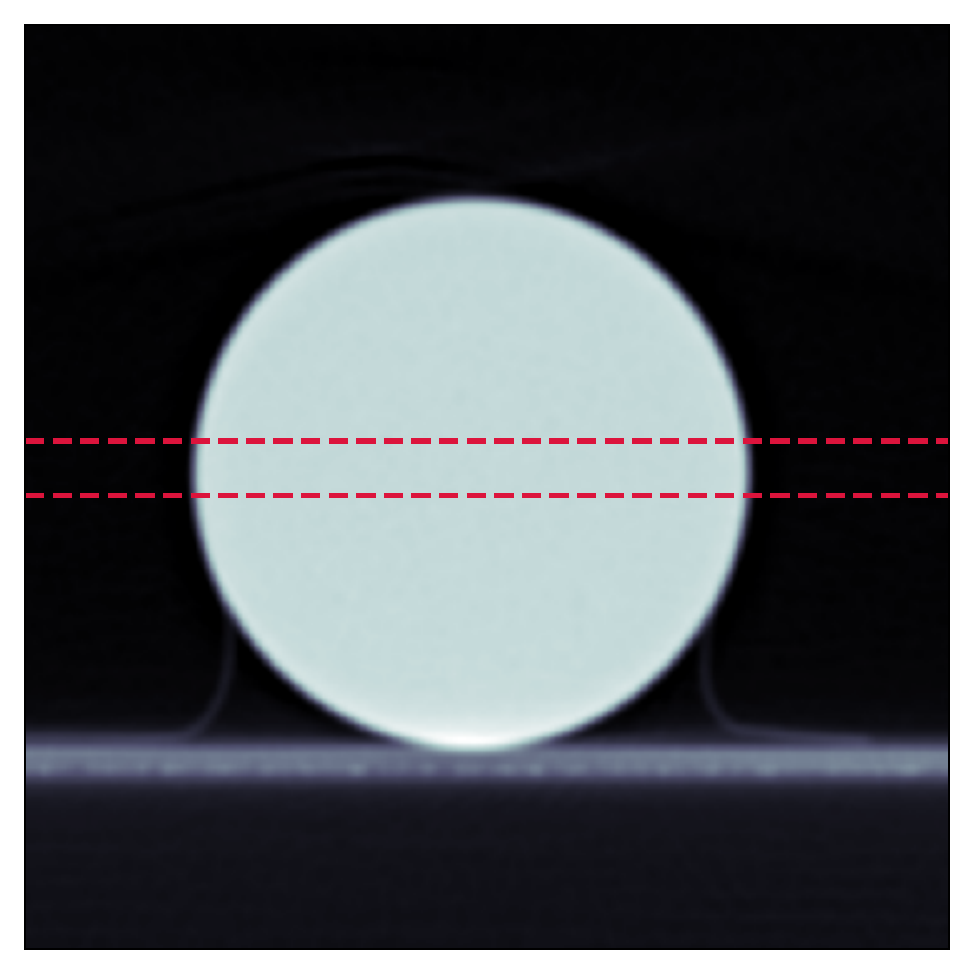}
    \caption{}
    \end{subfigure}
     \begin{subfigure}{.45\textwidth}
    \centering
    \includegraphics[width =0.8\textwidth ]{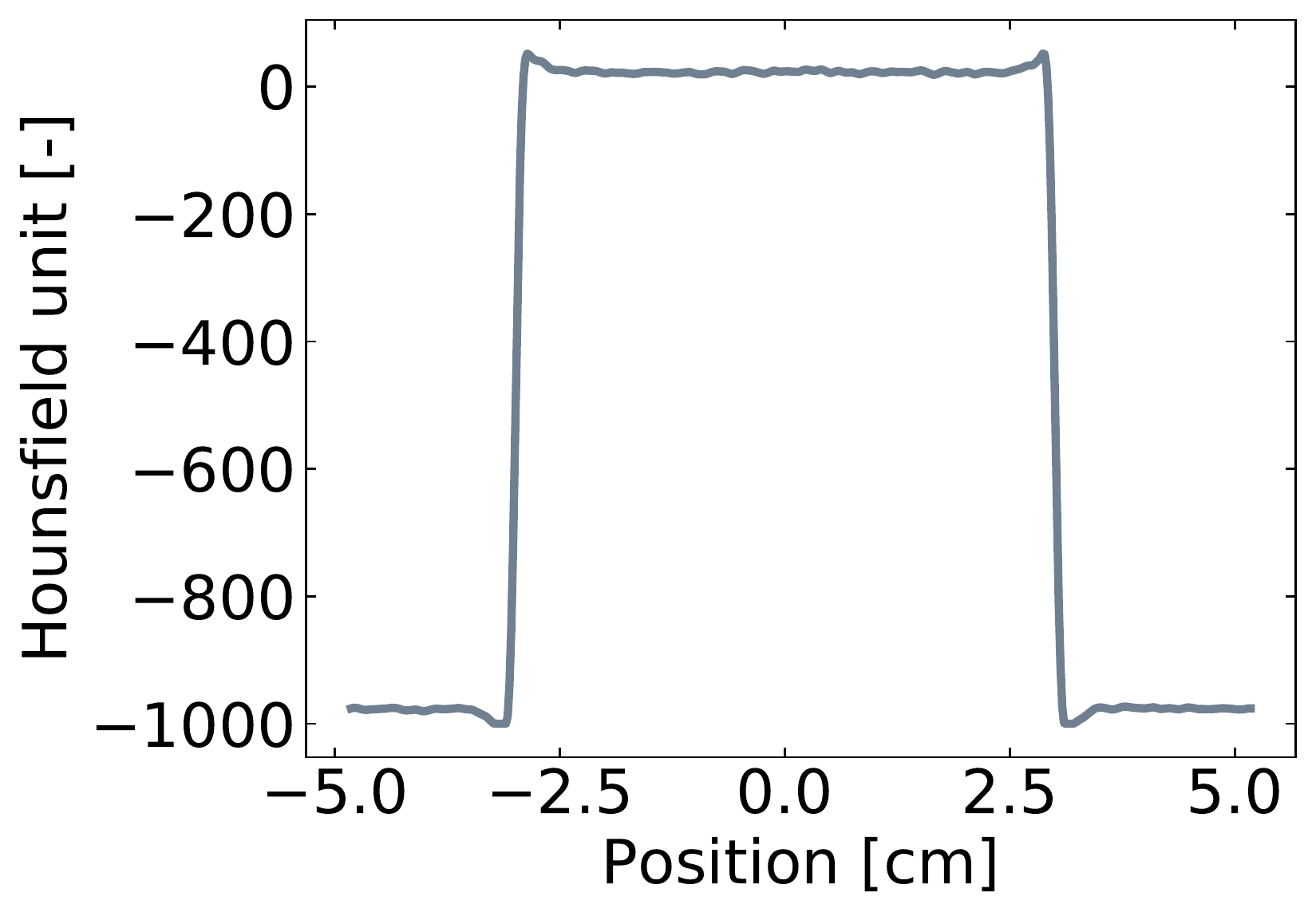}
    \caption{}
    \end{subfigure}
    \caption{Image of a CT slice of the dosimeter (a) and a Housfield Unit profile extracted from a region of interest (b). Dashed red lines on (a) correspond to the selected region of interest. }
    \label{fig:geldensity}
\end{figure}

Evaluation of the voxel density values from CT-scans yielded (mean $\pm$ standard deviation) densities of 1.002 $\pm$ 0.005, 1.000 $\pm$ 0.005 and 0.999 $\pm$ 0.005 g/cm$^3$ respectively for water, the urethane elastomer, and the elastomer with the scintillating fibers inside. Figure \ref{fig:geldensity} further presents a slice acquired from the CT and a profile drawn across a region of interest. Even if the region of interest intercepts four scintillating fibers, those are indistinguishable from the bulk elastomer.

\begin{figure}[ht]
    \centering
    \begin{subfigure}{.49\textwidth}
    \centering
    \includegraphics[width =0.8\textwidth ]{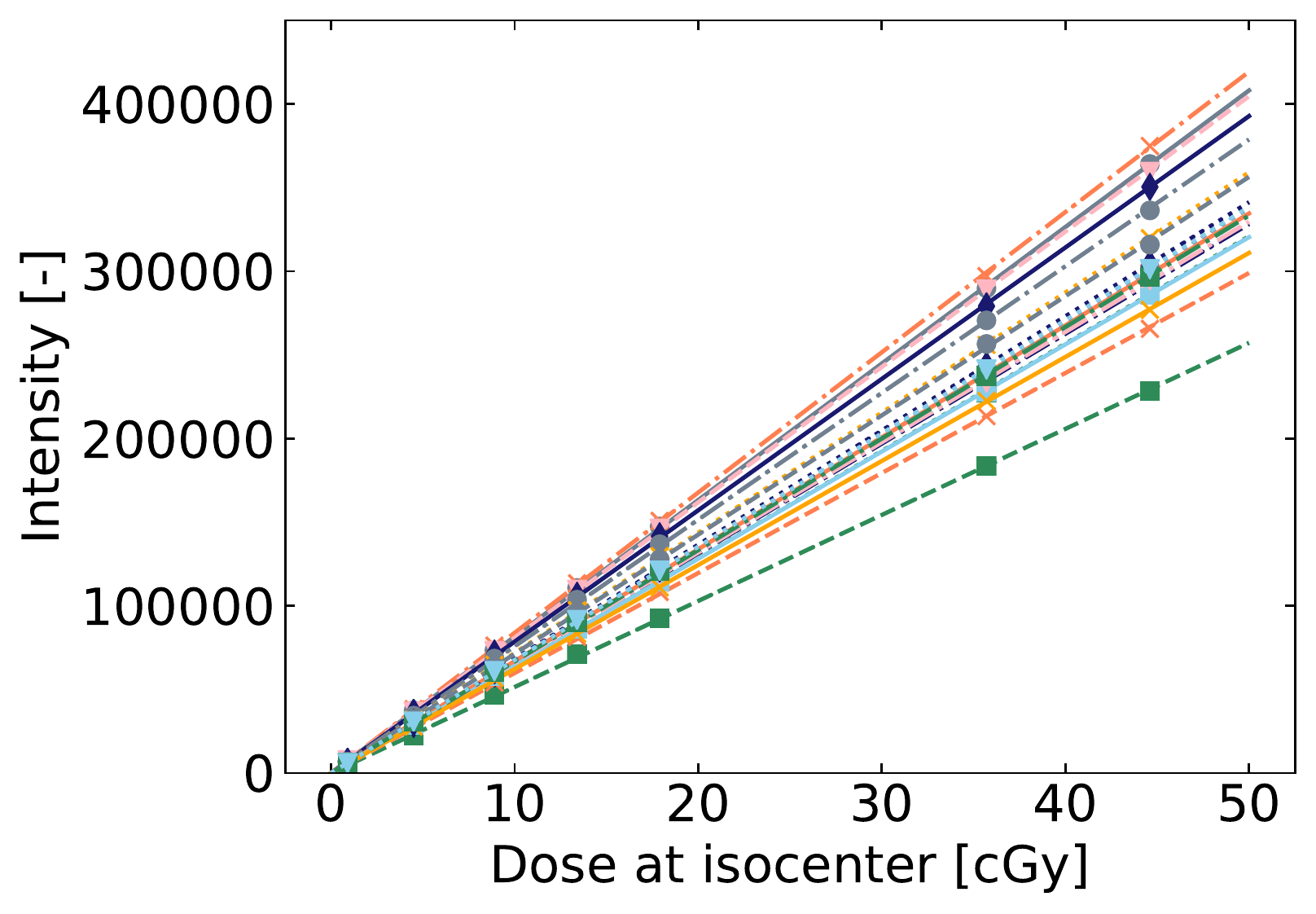}
    \caption{}
    \end{subfigure}
     \begin{subfigure}{.49\textwidth}
    \centering
    \includegraphics[width =0.8\textwidth ]{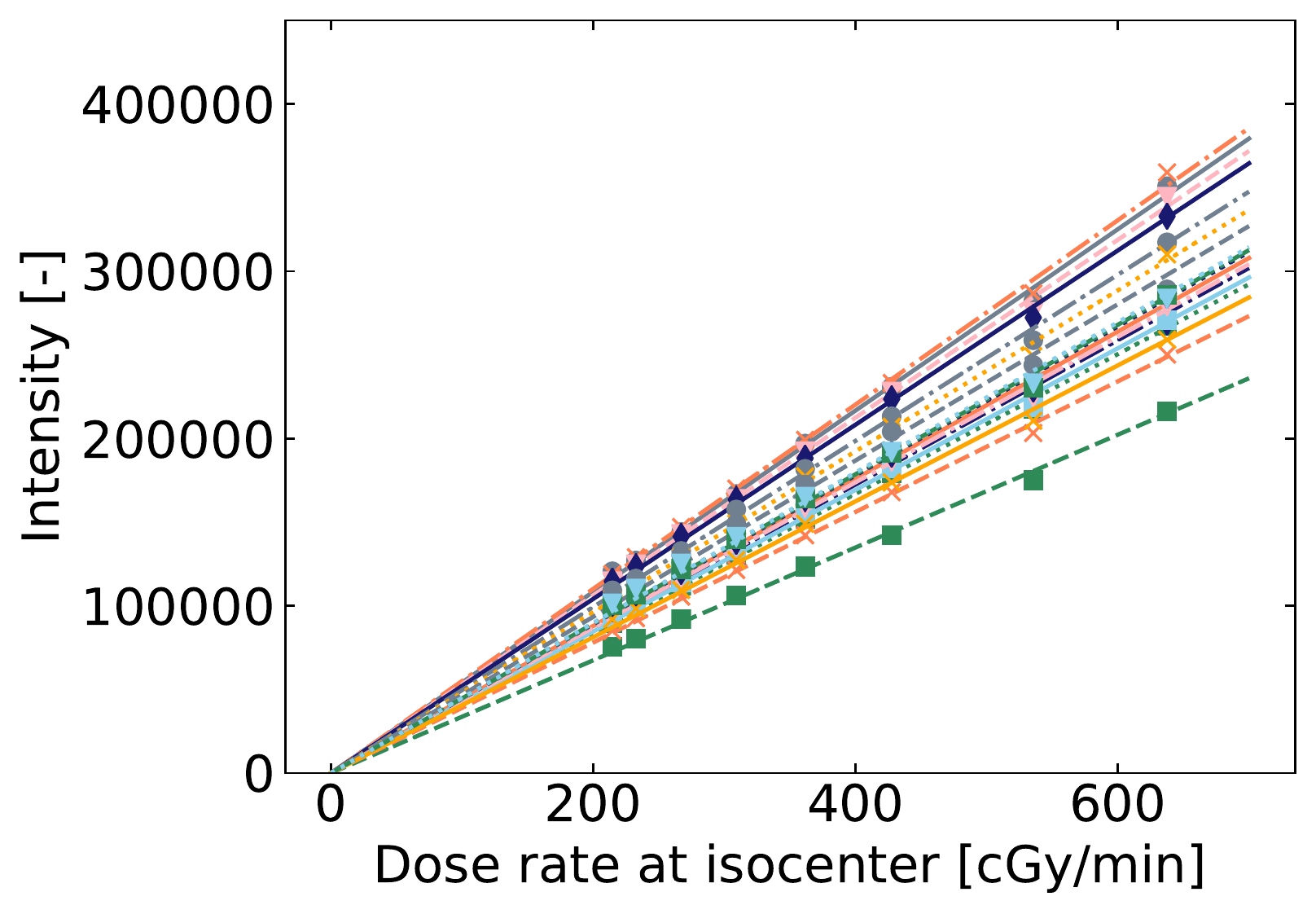}
    \caption{}
    \end{subfigure}
    \caption{Linearity of the scintillation signal as function of the dose (a) and dose rate (b) for all of the 19 scintillating fibers. Solid lines represent linear fits : $R^2>0.999$ for all measurements. }
    \label{fig:linearity}
\end{figure}

 As could be expected, the detector exhibited a linear dose-light relationship (R$^2$ $>$ 0.999) for all of the 19 scintillation fibers as presented on figure \ref{fig:linearity}(a). The signal to dose proportionality remained linear (R$^2$ $>$ 0.999) when varying the dose rate from 215 to 660 cGy/min (figure \ref{fig:linearity}(b)).

\subsubsection{Spatial dependencies correction}
\begin{figure}[ht]
    \centering
    \begin{subfigure}{.35\textwidth}
    \centering
    \includegraphics[width =0.9\textwidth ]{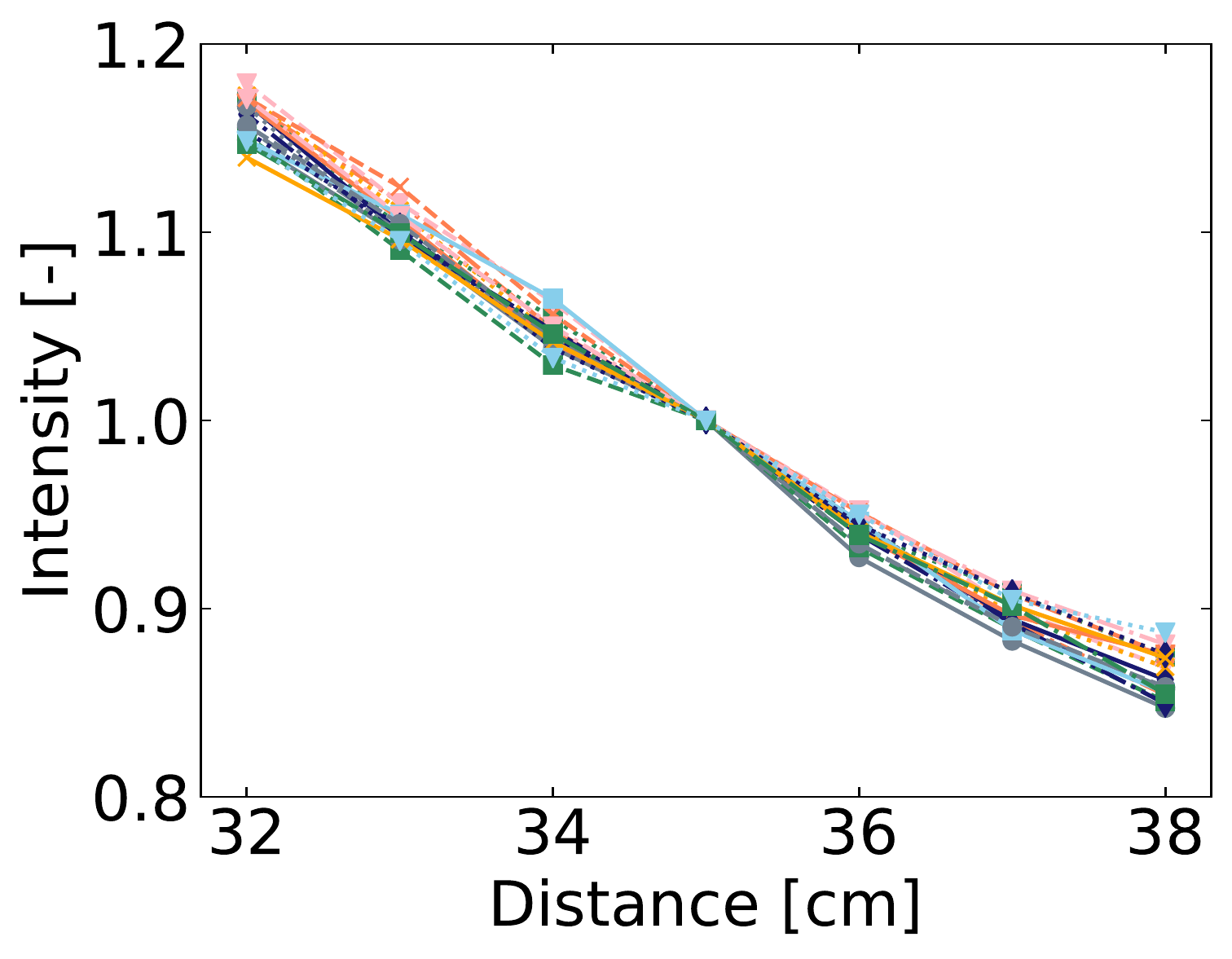}
    \caption{}
    \end{subfigure}
     \begin{subfigure}{.35\textwidth}
    \centering
    \includegraphics[width =0.9\textwidth ]{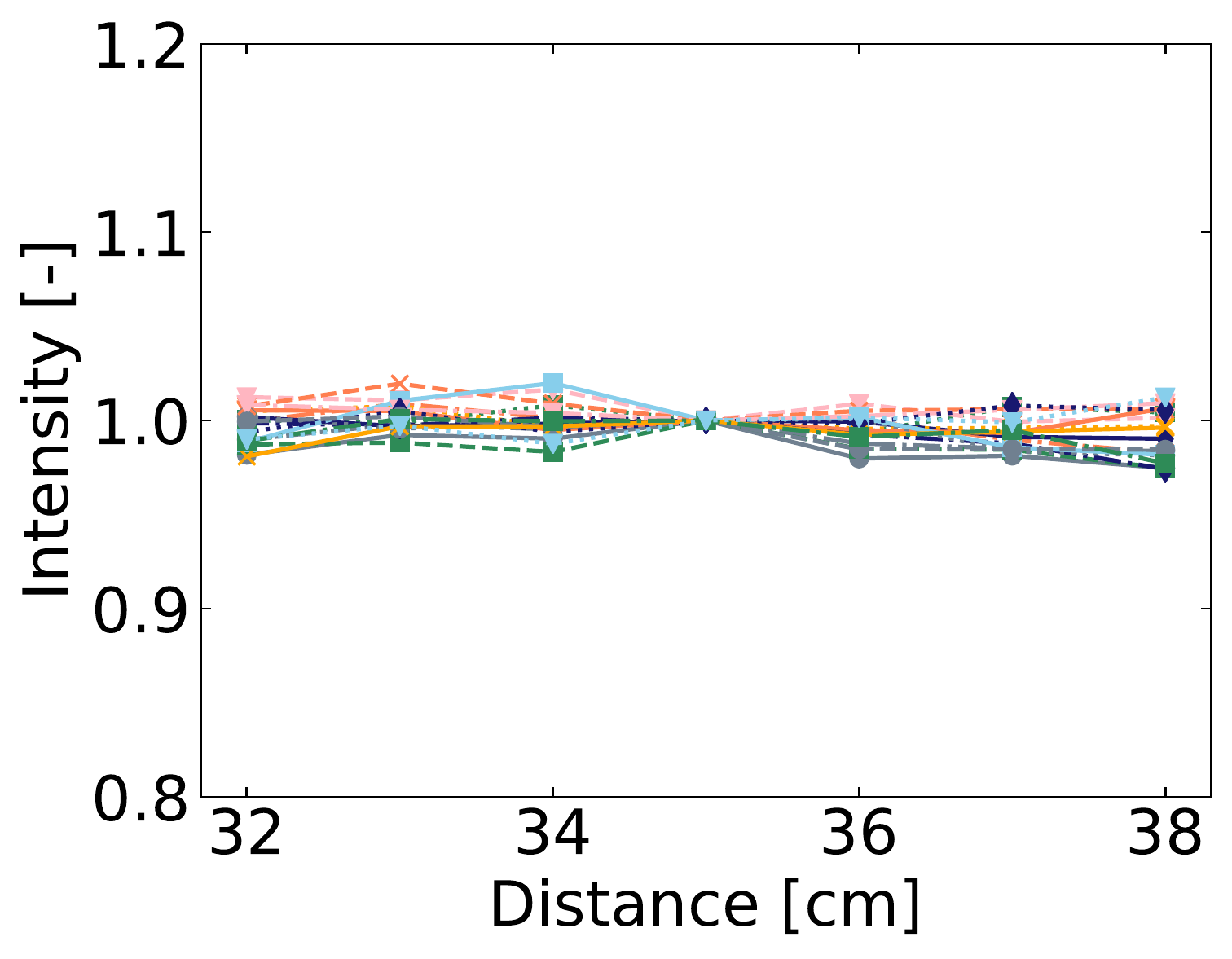}
    \caption{}
    \end{subfigure}
    \begin{subfigure}{.22\textwidth}
    \begin{subfigure}{\textwidth}
    \includegraphics[width =0.9\textwidth ]{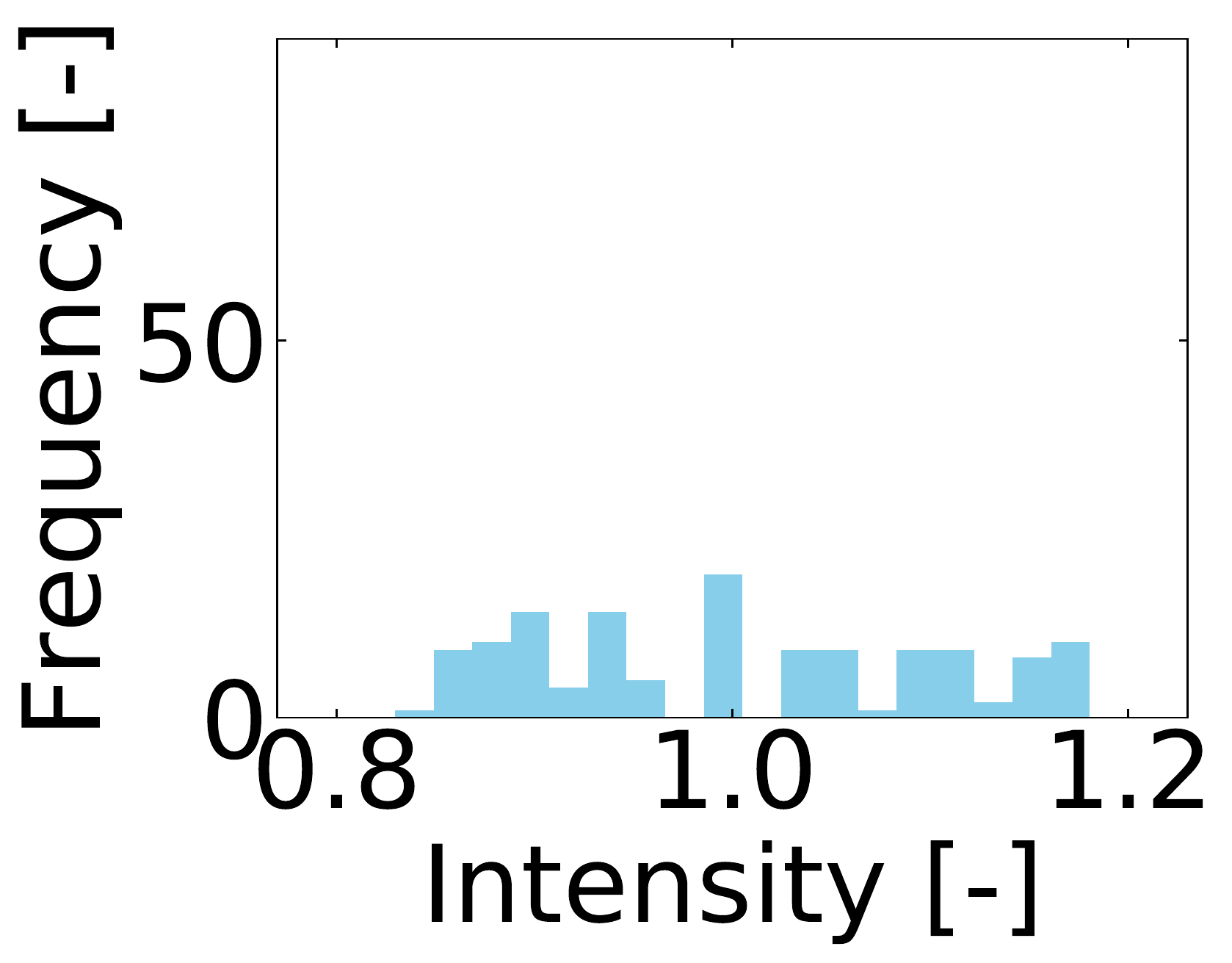}
    \caption{}
    \end{subfigure}
    \begin{subfigure}{\textwidth}
    \includegraphics[width =0.9\textwidth ]{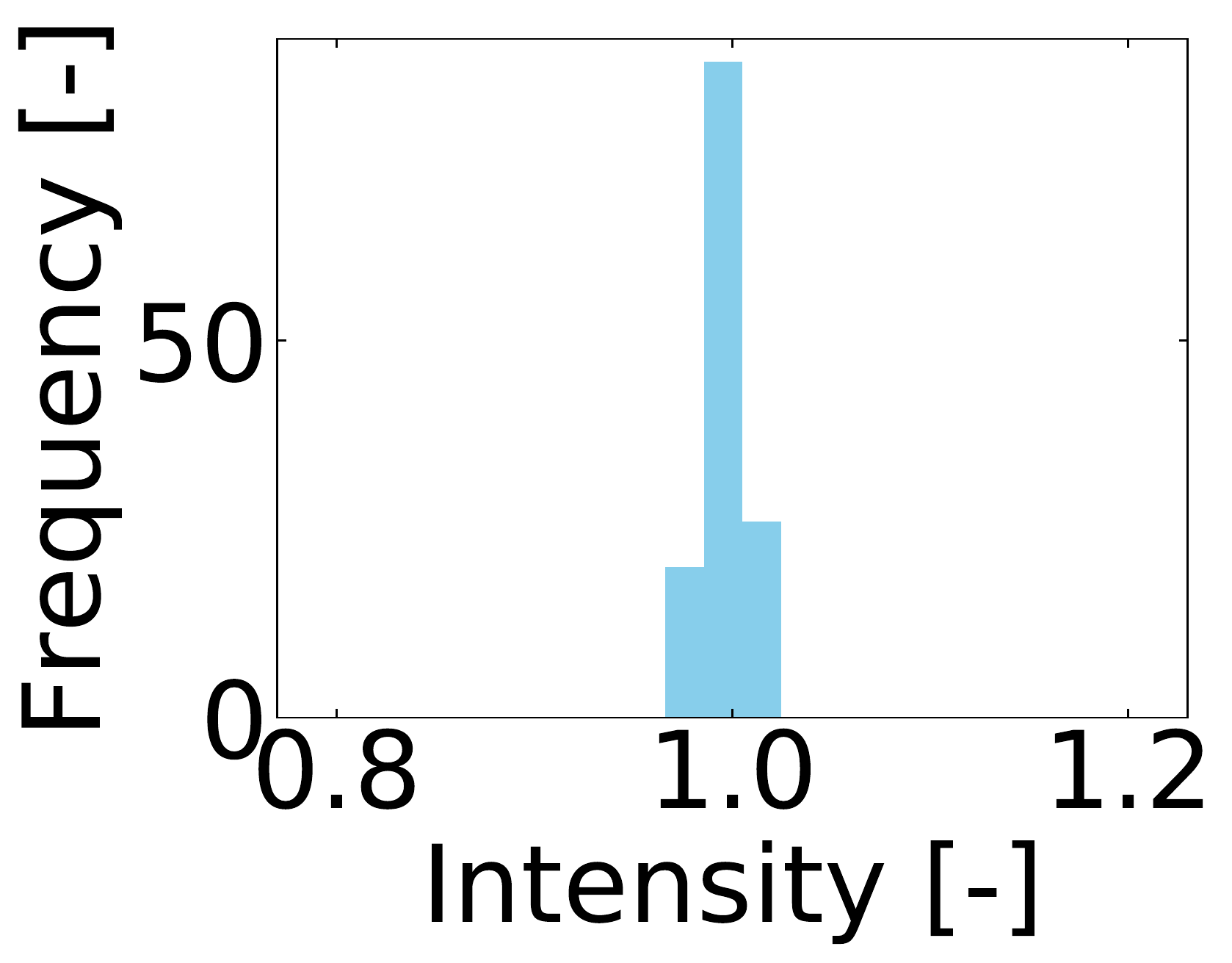}
    \caption{}
    \end{subfigure}
     
    \end{subfigure}
   
     \caption{Measured signal variation resulting from moving the dosimeter from a distance 32 to 38 cm in depth before (a) and after applying the signal corrections (b). (c) and (d) present the distribution of raw and corrected data, respectively. }
     \label{fig:corrdep}
\end{figure}

The signal obtained for the 19 scintillating fibers as a result of varying the distance between the gel and the CCD camera from 32 to 38 cm is presented of figure \ref{fig:corrdep}. The raw signal varies from 84.7\% to 117.9\% of the one obtained at 35 cm (used as reference). Applying the inverse square law using the 3D positioning of the fibers provided by the stereo matching cameras to the resulting signal reduced those variation to 97.4\% up to 101.9\%. Figures \ref{fig:corrdep}(c) et \ref{fig:corrdep}(d) present the distribution from all gathered data from the 19 scintillating fibers for each distance prior and after corrections are applied. Radial distance variations resulted in mean $\pm$ standard deviation intensities of 100$\pm$10 \% and 100$\pm$1\% before and after corrections, respectively. 

\begin{figure}[ht]
    \centering
    \begin{subfigure}{.35\textwidth}
    \centering
    \includegraphics[width =0.9\textwidth ]{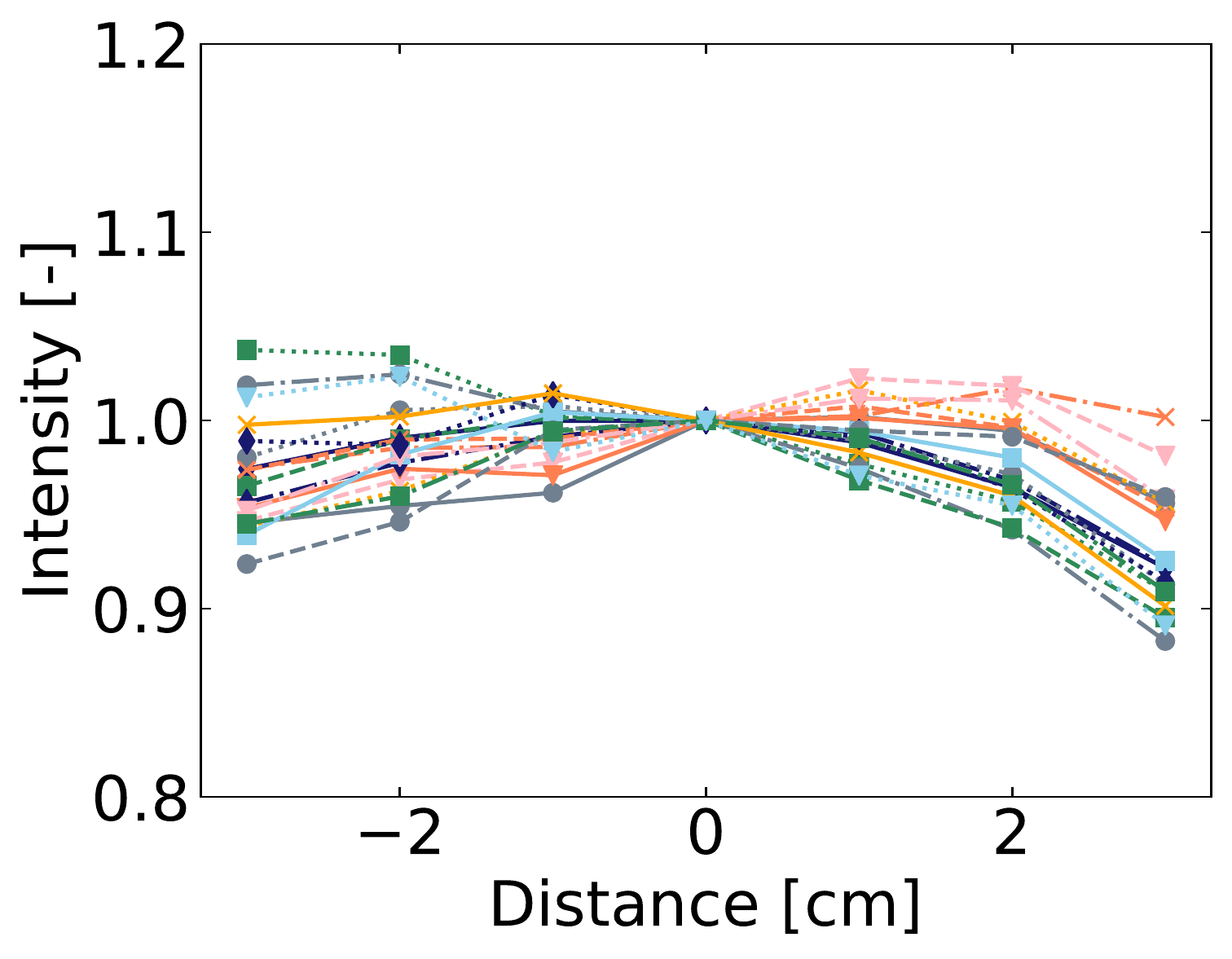}
    \caption{}
    \end{subfigure}
     \begin{subfigure}{.35\textwidth}
    \centering
    \includegraphics[width =0.9\textwidth ]{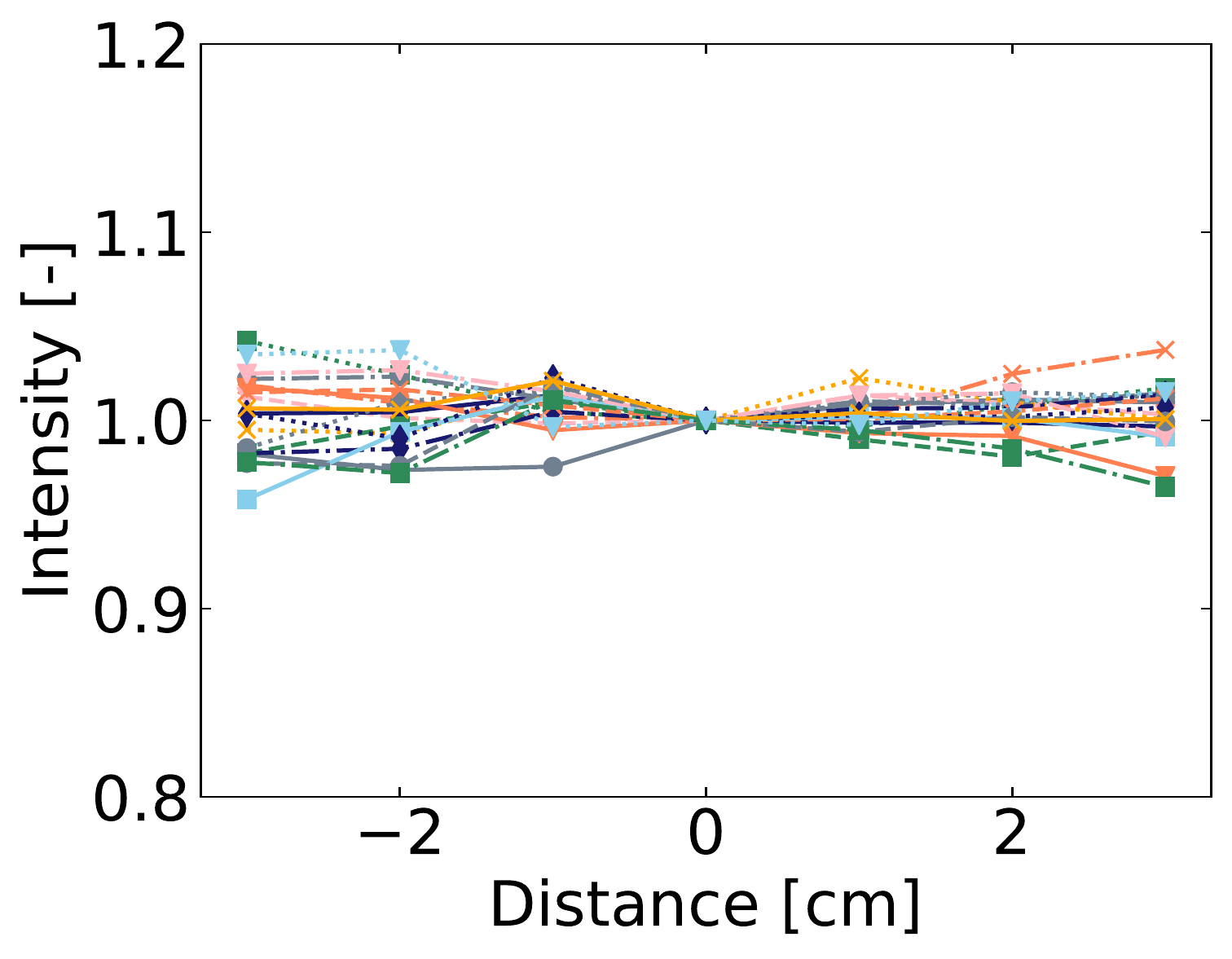}
    \caption{}
    \end{subfigure}
    \begin{subfigure}{.22\textwidth}
    \begin{subfigure}{\textwidth}
    \includegraphics[width =0.9\textwidth ]{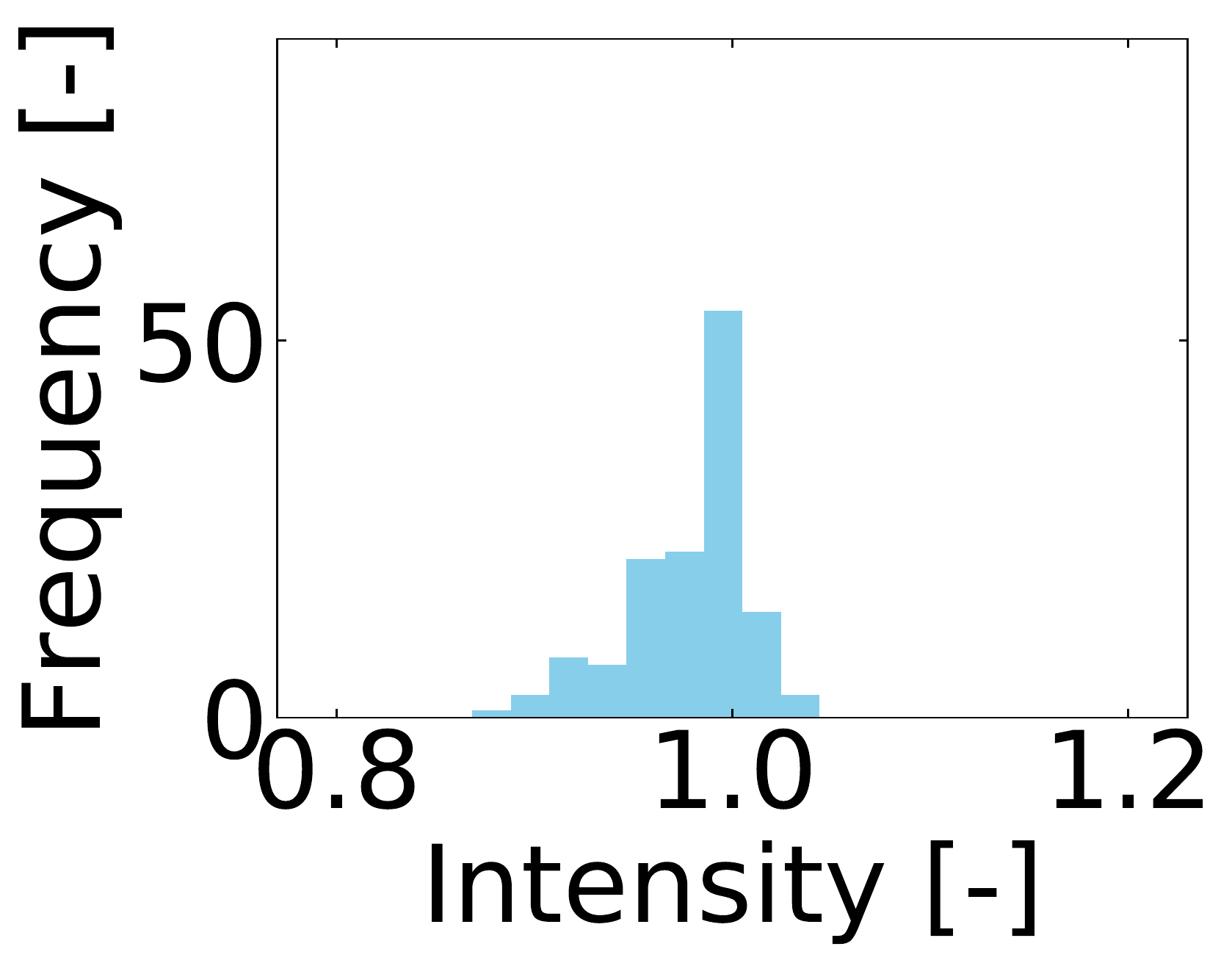}
    \caption{}
    \end{subfigure}
    \begin{subfigure}{\textwidth}
    \includegraphics[width =0.9\textwidth ]{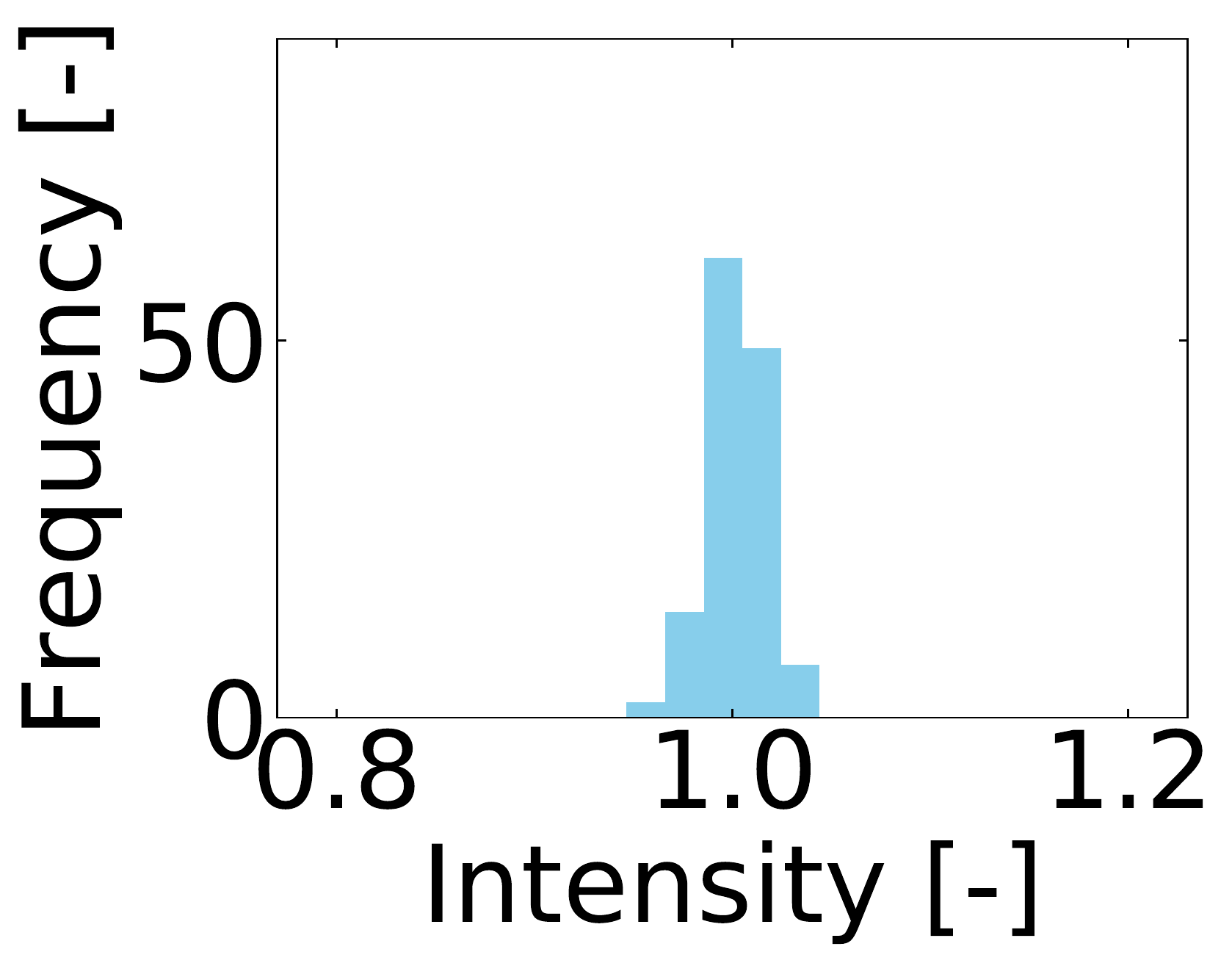}
    \caption{}
    \end{subfigure}
    
    \end{subfigure}
    
    \caption{Measured signal variation resulting from lateral displacement before (a) and after applying the signal corrections (b). (c) and (d) present the distribution of raw and corrected data, respectively.}
    \label{fig:corrlat}
\end{figure}

 Similarly, the raw signal variations caused by the lateral displacement of the deformable dosimeter are presented in figure 10. Displacing the dosimeter from -3 cm to 3 cm relative to it's initial position caused signal variations between 88.3\% and 103.7\%. Once corrected for angular and distal variations, signal variations ranged from 95.8\% to 104.2\%. The signal drop observed at 3 cm on figure \ref{fig:corrlat} (a) results from a small angulation of the prototype after its re-positioning at the isocenter. The angulation was detected by the system and corrected as seen on \ref{fig:corrlat} (b). The data distributions presented of figures \ref{fig:corrlat}(c) et \ref{fig:corrlat}(d) reveal mean $\pm$ standard deviation intensities of 98\%$\pm$3\% and 100\%$\pm$1\% before and after corrections, respectively.
 

\section{Discussion}

The signal produced by a scintillating fiber that is measured with a camera depends on the distance between the camera and the scintillating fiber as well as the angle between their respective axes. The resulting decrease of the measured signal as function of the tilt of the fiber arises from the combined gaussian output of the guided scintillation signal in the fiber and the non-guided isotropic signal emitted at the tip of the fiber. Signal variations related to the distance between the camera and the fibers follows the inverse square-law, as previously demonstrated \cite{bruza_time-gated_2018, xie_cherenkov_2020}. As a consequence, if not corrected, a deformation of 1 cm in the z axis, captured by a camera distant of 35 cm, would lead to signal variations of 5.48\%. Thus, tracking position and angulation of the fibers is essential for adequate dose measurements. 

This work proposes the use of computer vision techniques to track the position of scintillating fibers. The 3D optical position tracking enabled a precision of 0.03~cm. This is slightly larger than the tolerance on the robot's positioning  of 0.01 cm and the camera's pixel resolution of 0.02~cm. The discrepancy on figure \ref{fig:stereo}(a) happening when the fiber passes the robot's wrist center highlights the robot's singularity point (i.e. a configuration where the robot is blocked in certain direction, thus modifying its path). Overall, our prototype dosimeter constitutes an application well suited to stereo vision. Indeed, the accuracy of 3D reconstruction in stereo vision relies on 1) the feature detection, and 2) the feature matching. Solving the correspondence problem in the image pairs is one of the main challenges of stereo vision and many strategies have been proposed to solve it \cite{5606082}. Having 19 well-defined and organized points to match significantly eases that challenge. As a result, our uncertainties are limited to the feature detection, i.e. centroids, accuracy. Keeping the dosimetry application in mind, an uncertainty of 0.03 cm would lead to a 0.17\% dose uncertainty, at a distance of 35 cm. As for the angle measurement, the uncertainty of 2$\degree$ is consistent with the spatial resolution of the camera's limited to 0.02 cm. Compromising the field of view, with a longer focal length objective for example, would improve the spatial and resulting angular resolution of the system. The system could also be improved with the addition of another sCMOS forming a stereoscopic pair with the facing camera (sCMOS1) to completely position the fibers on both sides \cite{part2-submittedpaper}. 

Correction functions (distance, angle and vignetting) were validated before their application to the case of a deformable dosimeter. To do so, signal variations resulting from the motion of a sCMOS imaging an irradiated scintillating fiber, from -7 to 7 cm, were corrected using expected distances ($r$) and angles ($\theta$). Thus, a signal decrease down to 85\% was reduced to 0.5\%, after corrections. Using known $r$ and $\theta$ enabled a validation that minimized uncertainties related to the distance and angle measurement.


 Density of the deformable detector presented no significant difference with water, meaning the detector can simultaneously act as a water-equivalent detector and phantom. Overall, displacement of the dosimeter radially generated higher signal variations than lateral displacement. However, radial signal variations were more effectively accounted for as the extrema were brought closer to the reference in comparison to data resulting from lateral displacement. This highlights the accuracy of the 3D positioning provided by the stereoscopic pair that enabled an efficient distal correction. On the other hand, angular corrections rely on the measurements of small pixel shifts. As lateral displacement has a stronger angular than radial correction dependency, the corrected results exhibited larger discrepancies. Thus, the signal variations remaining in the corrected data are attributed to uncertainties on the angular measurement, mainly, and the 3D tracking accuracy. Smaller variations would also have to be expect from fiber to fiber variations, in particular from the polishing of their ends and exact lengths, which could affect the gaussian's FWHM which guides the angular correction. 

The use of sCMOS cameras in the set-up enabled the 3D positionning and orientation measurement of scintillating fibers. These cameras were used to acquire qualitative images of high spatial resolution. Hence, the accuracy of the 3D tracking and angle measurement only rely on the spatial resolution of the cameras, rather than their stability over time.  Thus, they could be replaced with cheaper cameras having a sufficient spatial resolution. Since the radiometry is solely carried by the CCD, the method does not compromise the stability and accuracy of dose measurements. Moreover, the use of cameras directly on the treatment couch to measure the light emitted by the scintillation fibers allows a cable-less setup. Hence, the detection system does not impact deformation  movement of the dosimeter. 

Previous studies have looked into the characterization of optical artifacts inherent to the use of plastic scintillators in static geometries \cite{robertson_optical_2014}. As such, corrections related to CCD, lens and scintillator tank artifacts were proven essentials to the resulting dose measurement accuracy. However, new challenges arise from changing the scintillator to detector geometry. Thus, tracking the position and orientation of radioluminescent elements has applications to free-space measurements by a camera where corrections are proven beneficial. In fact, scintillators have previously been used in set-ups where the distance to the camera can vary, for example from the motion of the camera or the breathing movement of a patient  \cite{tendler_characterization_2019, jenkins_automating_2016, bruza_time-gated_2018}. In the latter, the authors did not correct the signal for distance variations, but found that a distance uncertainty margin of 2 cm was required to keep the dose uncertainty under 1\%, for a camera placed at a distance of 400 cm. More recently, inverse-square distance corrections have been proposed to account for Cerenkov-based dose measurements using a large field of view \cite{xie_cherenkov_2020}. The distal correction required reached $\approx$ 10\% as the field of view covered a patient's whole body.  In this work, the stereo matching provides an uncertainty margin of 0.03~cm, thus reducing the potential dose uncertainty or even allowing the positioning of the scintillator closer to the cameras. Therefore, precise 3D positioning and the associated corrections can increase the signal-to-noise ratio (SNR) of dose measurements using radioluminescent elements. Distance and angular dependencies corrections will be essential to accurate dose measurements when extending scintillator's range of application to deformable cases: deforming a phantom comprising scintillators will lead to translations and rotations of the scintillating elements. Translations up to 3 cm in every directions cover the expected range of displacements in the context of radiotherapy treatments. Indeed, anatomical variations in the millimeter-centimeter range occur through the course of treatments \cite{yeo_novel_2012}. A deformable scintillator-based detector would enable water-equivalent, real-time dose measurement to study the dosimetric impact of anatomical deformations. 

Compared to previously proposed scintillating dosimetry systems, this work enables dose measurements of a moving and deforming scintillation based dosimeter. Moreover, since scintillating fibers are read in free space, i.e. with no optical guide connecting them to a detection device, their position can be optically tracked and serve as markers. Hence, the optical 3D position tracking can be use to measure positions and resulting deformation. Future work will look into the dosimetric characterisation of the system as well as its ability to measure deformation vector fields. A water-equivalent, real-time dosimeter that simultaneously measures dose and the deformation vector field would have extensive application in validating deformable image registration algorithms \cite{kirby_need_2013, kirby_two-dimensional_2011} as well as understanding the dosimetric impact of anatomical deformations.

\section{Conclusion}
The novel use of scintillating fibers in a varying geometry phantom presented new difficulties that were characterized and corrected. Hence, measurements of the angular and distal variations of the fibers from the detector reduced the signal dependencies on the varying geometry of the gel. Pairing a cooled CCD to two sCMOS enabled the 3D positionning and angular tracking of 19 moving scintillating fibers. All together, the setup enabled a correction workflow accounting for distal and angular variations of moving scintillating elements. Moreover, we prototyped a novel deformable scintillation detector measuring the dose at 19 points in a flexible phantom. This works is a step toward the use of plastic scintillators in moving and varying geometries.

\section{Acknowledgement}
The authors thank Serge Groleau for his help manufacturing the elastomer. This work was financed by the Natural Sciences and Engineering Research Council of Canada (NSERC) Discovery grants \#2019-05038 and \#2018-04055. Emily Cloutier acknowledges support by the Fonds de Recherche du Quebec – Nature et Technologies (FRQNT). The authors thank Ghyslain Leclerc for the English revision of the paper.



\begin{small}
\printbibliography[heading=bibintoc]

@article{part2-submittedpaper, 
Author = {Cloutier, Emily and Beaulieu, Luc and Archambault, Louis},
Title = {Deformable Scintillation Dosimeter II: Real-Time Simultaneous Measurements of Dose and Tracking of Deformation Vector Fields},
Date = {2021}
}

@article{madden_first_2019,
	title = {First measurements with a plastic scintillation dosimeter at the Australian {MRI}-{LINAC}},
	volume = {64},
	issn = {1361-6560},
	url = {https://iopscience.iop.org/article/10.1088/1361-6560/ab324b},
	doi = {10.1088/1361-6560/ab324b},
	pages = {175015},
	number = {17},
	journaltitle = {Physics in Medicine \& Biology},
	shortjournal = {Phys. Med. Biol.},
	author = {Madden, Levi and Archer, James and Li, Enbang and Jelen, Urszula and Dong, Bin and Roberts, Natalia and Holloway, Lois and Rosenfeld, Anatoly},
	urldate = {2020-09-02},
	date = {2019-09-04},
	langid = {english}
}

@article{rilling_tomographicbased_2020,
	title = {Tomographic‐based {3D} scintillation dosimetry using a three‐view plenoptic imaging system},
	issn = {0094-2405, 2473-4209},
	url = {https://onlinelibrary.wiley.com/doi/abs/10.1002/mp.14213},
	doi = {10.1002/mp.14213},
	language = {en},
	urldate = {2020-05-28},
	journal = {Medical Physics},
	author = {Rilling, Madison and Allain, Guillaume and Thibault, Simon and Archambault, Louis},
	month = may,
	year = {2020},
	pages = {mp.14213}
}

@article{thompson_artificial_2018,
	title = {Artificial intelligence in radiation oncology: {A} specialty-wide disruptive transformation?},
	volume = {129},
	issn = {01678140},
	shorttitle = {Artificial intelligence in radiation oncology},
	url = {https://linkinghub.elsevier.com/retrieve/pii/S0167814018302895},
	doi = {10.1016/j.radonc.2018.05.030},
	language = {en},
	number = {3},
	urldate = {2020-08-06},
	journal = {Radiotherapy and Oncology},
	author = {Thompson, Reid F. and Valdes, Gilmer and Fuller, Clifton D. and Carpenter, Colin M. and Morin, Olivier and Aneja, Sanjay and Lindsay, William D. and Aerts, Hugo J.W.L. and Agrimson, Barbara and Deville, Curtiland and Rosenthal, Seth A. and Yu, James B. and Thomas, Charles R.},
	month = dec,
	year = {2018},
	pages = {421--426}
}

@INPROCEEDINGS{5606082, author={M. {Gosta} and M. {Grgic}}, booktitle={Proceedings ELMAR-2010}, title={Accomplishments and challenges of computer stereo vision}, year={2010}, volume={}, number={}, pages={57-64},}

@article{guillot_performance_2013,
	title = {Performance assessment of a {2D} array of plastic scintillation detectors for {IMRT} quality assurance},
	volume = {58},
	issn = {0031-9155, 1361-6560},
	url = {https://iopscience.iop.org/article/10.1088/0031-9155/58/13/4439},
	doi = {10.1088/0031-9155/58/13/4439},
	language = {en},
	number = {13},
	urldate = {2020-04-24},
	journal = {Physics in Medicine and Biology},
	author = {Guillot, Mathieu and Gingras, Luc and Archambault, Louis and Beddar, Sam and Beaulieu, Luc},
	month = jul,
	year = {2013},
	pages = {4439--4454}
}

@article{yeo_novel_2012,
	title = {A novel methodology for 3D deformable dosimetry},
	volume = {39},
	issn = {0094-2405},
	url = {http://scitation.aip.org.acces.bibl.ulaval.ca/content/aapm/journal/medphys/39/4/10.1118/1.3694107},
	doi = {10.1118/1.3694107},
	pages = {2203--2213},
	number = {4},
	journaltitle = {Medical Physics},
	author = {Yeo, U. J. and Taylor, M. L. and Dunn, L. and Kron, T. and Smith, R. L. and Franich, R. D.},
	urldate = {2016-05-11},
	date = {2012-04-01}
}

@article{kirov_three-dimensional_2005,
	title = {The three-dimensional scintillation dosimetry method: test for a 106 Ru eye plaque applicator},
	volume = {50},
	issn = {0031-9155},
	url = {http://stacks.iop.org/0031-9155/50/i=13/a=007},
	doi = {10.1088/0031-9155/50/13/007},
	shorttitle = {The three-dimensional scintillation dosimetry method},
	pages = {3063},
	number = {13},
	journaltitle = {Physics in Medicine and Biology},
	shortjournal = {Phys. Med. Biol.},
	author = {Kirov, A. S. and Piao, J. Z. and Mathur, N. K. and Miller, T. R. and Devic, S. and Trichter, S. and Zaider, M. and Soares, C. G. and {LoSasso}, T.},
	urldate = {2016-05-17},
	date = {2005},
	langid = {english}
}

@article{goulet_novel_2014,
	title = {Novel, full 3D scintillation dosimetry using a static plenoptic camera},
	volume = {41},
	issn = {0094-2405},
	url = {http://scitation.aip.org.acces.bibl.ulaval.ca/content/aapm/journal/medphys/41/8/10.1118/1.4884036},
	doi = {10.1118/1.4884036},
	pages = {082101},
	number = {8},
	journaltitle = {Medical Physics},
	author = {Goulet, Mathieu and Rilling, Madison and Gingras, Luc and Beddar, Sam and Beaulieu, Luc and Archambault, Louis},
	urldate = {2016-05-31},
	date = {2014-08-01}
}

@article{kirov_new_2000,
	title = {New water equivalent liquid scintillation solutions for 3D dosimetry},
	volume = {27},
	issn = {0094-2405},
	url = {http://scitation.aip.org.acces.bibl.ulaval.ca/content/aapm/journal/medphys/27/5/10.1118/1.598993},
	doi = {10.1118/1.598993},
	pages = {1156--1164},
	number = {5},
	journaltitle = {Medical Physics},
	author = {Kirov, A. S. and Shrinivas, S. and Hurlbut, C. and Dempsey, J. F. and Binns, W. R. and Poblete, J. L.},
	urldate = {2016-05-16},
	date = {2000-05-01}
}

@article{schaly_tracking_2004,
	title = {Tracking the dose distribution in radiation therapy by accounting for variable anatomy},
	volume = {49},
	issn = {0031-9155},
	url = {http://stacks.iop.org/0031-9155/49/i=5/a=010},
	doi = {10.1088/0031-9155/49/5/010},
	pages = {791},
	number = {5},
	journaltitle = {Physics in Medicine and Biology},
	shortjournal = {Phys. Med. Biol.},
	author = {Schaly, B. and Kempe, J. A. and Bauman, G. S. and Battista, J. J. and Dyk, J. Van},
	urldate = {2016-05-25},
	date = {2004},
	langid = {english}
}

@article{kroll_preliminary_2013,
	title = {Preliminary investigations on the determination of three-dimensional dose distributions using scintillator blocks and optical tomography},
	volume = {40},
	issn = {0094-2405},
	url = {http://scitation.aip.org.acces.bibl.ulaval.ca/content/aapm/journal/medphys/40/8/10.1118/1.4813898},
	doi = {10.1118/1.4813898},
	pages = {082104},
	number = {8},
	journaltitle = {Medical Physics},
	author = {Kroll, Florian and Pawelke, Jörg and Karsch, Leonhard},
	urldate = {2016-07-19},
	date = {2013-08-01}
}

@article{beaulieu_review_2016,
	title = {Review of plastic and liquid scintillation dosimetry for photon, electron, and proton therapy},
	volume = {61},
	issn = {0031-9155, 1361-6560},
	url = {http://stacks.iop.org/0031-9155/61/i=20/a=R305?key=crossref.88afc16233fc2bff5a63aa33e763bd8d},
	doi = {10.1088/0031-9155/61/20/R305},
	pages = {R305--R343},
	number = {20},
	journaltitle = {Physics in Medicine and Biology},
	author = {Beaulieu, Luc and Beddar, Sam},
	urldate = {2016-10-07},
	date = {2016-10-21}
}

@article{beddar_water_2006,
	title = {Water equivalent plastic scintillation detectors in radiation therapy},
	volume = {120},
	issn = {0144-8420, 1742-3406},
	url = {http://rpd.oxfordjournals.org.acces.bibl.ulaval.ca/content/120/1-4/1},
	doi = {10.1093/rpd/nci694},
	pages = {1--6},
	number = {1},
	journaltitle = {Radiation Protection Dosimetry},
	shortjournal = {Radiat Prot Dosimetry},
	author = {Beddar, A. S.},
	urldate = {2016-11-21},
	date = {2006-01-09},
	langid = {english},
	pmid = {16882685}
}

@book{beddar_scintillation_2016,
	title = {Scintillation Dosimetry},
	isbn = {978-1-4822-0899-3},
	url = {https://www.crcpress.com/Scintillation-Dosimetry/Beddar-Beaulieu/p/book/9781482208993},
	publisher = {{CRC} Press},
	author = {Beddar, A. S. and Beaulieu, L},
	urldate = {2016-11-25},
	date = {2016-04-01}
}

@article{kirby_need_2013,
	title = {The need for application-based adaptation of deformable image registration},
	volume = {40},
	number = {1},
	journaltitle = {Medical physics},
	author = {Kirby, Neil and Chuang, Cynthia and Ueda, Utako and Pouliot, Jean},
	date = {2013}
}

@article{kirby_two-dimensional_2011,
	title = {A two-dimensional deformable phantom for quantitatively verifying deformation algorithms},
	volume = {38},
	issn = {2473-4209},
	url = {http://onlinelibrary.wiley.com.acces.bibl.ulaval.ca/doi/10.1118/1.3597881/abstract},
	doi = {10.1118/1.3597881},
	pages = {4583--4586},
	number = {8},
	journaltitle = {Medical Physics},
	shortjournal = {Med. Phys.},
	author = {Kirby, Neil and Chuang, Cynthia and Pouliot, Jean},
	urldate = {2017-11-16},
	date = {2011-08-01},
	langid = {english}
}

@article{pogue_cherenkov_2015,
	title = {Cherenkov radiation dosimetry in water tanks – video rate imaging, tomography and {IMRT} \& {VMAT} plan verification},
	volume = {573},
	issn = {1742-6596},
	url = {http://stacks.iop.org/1742-6596/573/i=1/a=012013},
	doi = {10.1088/1742-6596/573/1/012013},
	pages = {012013},
	number = {1},
	journaltitle = {Journal of Physics: Conference Series},
	shortjournal = {J. Phys.: Conf. Ser.},
	author = {Pogue, Brian W. and Glaser, Adam K. and Zhang, Rongxiao and Gladstone, David J.},
	urldate = {2017-11-20},
	date = {2015},
	langid = {english}
}

@article{archambault_transient_2008,
	title = {Transient noise characterization and filtration in {CCD} cameras exposed to stray radiation from a medical linear accelerator},
	volume = {35},
	issn = {2473-4209},
	url = {http://onlinelibrary.wiley.com.acces.bibl.ulaval.ca/doi/10.1118/1.2975147/abstract},
	doi = {10.1118/1.2975147},
	pages = {4342--4351},
	number = {10},
	journaltitle = {Medical Physics},
	shortjournal = {Med. Phys.},
	author = {Archambault, Louis and Briere, Tina Marie and Beddar, Sam},
	urldate = {2017-11-29},
	date = {2008-10-01},
	langid = {english}
}

@article{robertson_optical_2014,
	title = {Optical artefact characterization and correction in volumetric scintillation dosimetry},
	volume = {59},
	issn = {0031-9155},
	url = {http://stacks.iop.org/0031-9155/59/i=1/a=23},
	doi = {10.1088/0031-9155/59/1/23},
	pages = {23},
	number = {1},
	journaltitle = {Physics in Medicine \& Biology},
	shortjournal = {Phys. Med. Biol.},
	author = {Robertson, Daniel and Hui, Cheukkai and Archambault, Louis and Mohan, Radhe and Beddar, Sam},
	urldate = {2017-12-11},
	date = {2014},
	langid = {english}
}

@article{bruza_time-gated_2018,
	title = {Time-gated scintillator imaging for real-time optical surface dosimetry in total skin electron therapy},
	volume = {63},
	issn = {1361-6560},
	url = {http://stacks.iop.org/0031-9155/63/i=9/a=095009?key=crossref.b5e7dfba567b8f80e169c22e79552bf5},
	doi = {10.1088/1361-6560/aaba19},
	pages = {095009},
	number = {9},
	journaltitle = {Physics in Medicine \& Biology},
	shortjournal = {Phys. Med. Biol.},
	author = {Bruza, Petr and Gollub, Sarah L and Andreozzi, Jacqueline M and Tendler, Irwin I and Williams, Benjamin B and Jarvis, Lesley A and Gladstone, David J and Pogue, Brian W},
	urldate = {2019-06-05},
	date = {2018-05-02},
	langid = {english}
}

@article{rilling_simulating_nodate,
	title = {Simulating imaging-based tomographic systems using an optical design software for resolving 3D structures of translucent media},
	pages = {11},
	journaltitle = {Applied Optics},
	author = {Rilling, Madison and Archambault, Louis and Thibault, Simon},
	langid = {english}
}

@article{alexander_1_nodate,
	title = {1 Scintillation Imaging as a High-Resolution, Remote, Versatile 2D Detection 2 System for {MR}-Linac Quality Assurance},
	pages = {24},
	author = {Alexander, Daniel A and Zhang, Rongxiao and Bruza, Petr and Pogue, Brian W},
	langid = {english}
}

@article{jenkins_automating_2016,
	title = {Automating quality assurance of digital linear accelerators using a radioluminescent phosphor coated phantom and optical imaging},
	volume = {61},
	issn = {0031-9155, 1361-6560},
	url = {https://iopscience.iop.org/article/10.1088/0031-9155/61/17/L29},
	doi = {10.1088/0031-9155/61/17/L29},
	pages = {L29--L37},
	number = {17},
	journaltitle = {Physics in Medicine and Biology},
	shortjournal = {Phys. Med. Biol.},
	author = {Jenkins, Cesare H and Naczynski, Dominik J and Yu, Shu-Jung S and Yang, Yong and Xing, Lei},
	urldate = {2020-04-20},
	date = {2016-09-07},
	langid = {english}
}

@book{hartley_multiple_2004,
	title = {Multiple view geometry in computer vision},
	isbn = {978-0-511-18711-7},
	url = {https://doi.org/10.1017/CBO9780511811685},
	author = {Hartley, Richard and Zisserman, Andrew},
	urldate = {2020-04-21},
	date = {2004},
	langid = {english},
	note = {{OCLC}: 171123855}
}

@article{zhang_flexible_2000,
	title = {A flexible new technique for camera calibration},
	volume = {22},
	issn = {01628828},
	url = {http://ieeexplore.ieee.org/document/888718/},
	doi = {10.1109/34.888718},
	pages = {1330--1334},
	number = {11},
	journaltitle = {{IEEE} Transactions on Pattern Analysis and Machine Intelligence},
	shortjournal = {{IEEE} Trans. Pattern Anal. Machine Intell.},
	author = {Zhang, Z.},
	urldate = {2020-04-21},
	date = {2000-11},
	langid = {english}
}

@article{therriault-proulx_effect_2018,
	title = {Effect of magnetic field strength on plastic scintillation detector response},
	volume = {116},
	issn = {13504487},
	url = {https://linkinghub.elsevier.com/retrieve/pii/S1350448717308491},
	doi = {10.1016/j.radmeas.2018.06.011},
	pages = {10--13},
	journaltitle = {Radiation Measurements},
	shortjournal = {Radiation Measurements},
	author = {Therriault-Proulx, F. and Wen, Z. and Ibbott, G. and Beddar, S.},
	urldate = {2020-04-21},
	date = {2018-09},
	langid = {english}
}

@article{brock_inclusion_2003,
	title = {Inclusion of organ deformation in dose calculations},
	volume = {30},
	issn = {2473-4209},
	url = {http://onlinelibrary.wiley.com/doi/10.1118/1.1539039/abstract},
	doi = {10.1118/1.1539039},
	pages = {290--295},
	number = {3},
	journaltitle = {Medical Physics},
	shortjournal = {Med. Phys.},
	author = {Brock, K. K. and {McShan}, D. L. and Ten Haken, R. K. and Hollister, S. J. and Dawson, L. A. and Balter, J. M.},
	urldate = {2017-10-24},
	date = {2003-03-01},
	langid = {english}
}

@article{tendler_characterization_2019,
	title = {Characterization of a non-contact imaging scintillator-based dosimetry system for total skin electron therapy},
	volume = {64},
	issn = {1361-6560},
	url = {https://iopscience.iop.org/article/10.1088/1361-6560/ab1d8a},
	doi = {10.1088/1361-6560/ab1d8a},
	language = {en},
	number = {12},
	urldate = {2020-04-24},
	journal = {Physics in Medicine \& Biology},
	author = {Tendler, Irwin I and Bruza, Petr and Jermyn, Mike and Cao, Xu and Williams, Benjamin B and Jarvis, Lesley A and Pogue, Brian W and Gladstone, David J},
	month = jun,
	year = {2019},
	pages = {125025}
}

@article{opencv_library, 
    author = {Bradski, G.}, 
    journal = {Dr. Dobb's Journal of Software Tools}, 
    title = {{The OpenCV Library}}, 
    year = {2000} 
}

@article{xie_cherenkov_2020,
	title = {Cherenkov imaging for total skin electron therapy ({TSET})},
	volume = {47},
	issn = {0094-2405, 2473-4209},
	url = {https://onlinelibrary.wiley.com/doi/abs/10.1002/mp.13881},
	doi = {10.1002/mp.13881},
	language = {en},
	number = {1},
	urldate = {2020-07-12},
	journal = {Medical Physics},
	author = {Xie, Yunhe and Petroccia, Heather and Maity, Amit and Miao, Tianshun and Zhu, Yihua and Bruza, Petr and Pogue, Brian W. and Plastaras, John P. and Dong, Lei and Zhu, Timothy C.},
	month = jan,
	year = {2020},
	pages = {201--212}
}
\end{small}

\end{document}